\documentclass[a4paper,fleqn,usenatbib]{mnras}

\usepackage{newtxtext,newtxmath}

\usepackage[T1]{fontenc}
\usepackage{ae,aecompl}

\usepackage{graphicx}	
\usepackage{amsmath}	
\usepackage{amssymb}	

\title[Metallicity of Cepheids in the inner Disk]{First metallicity determination from Near-Infrared spectra for five obscured Cepheids discovered in the inner Disk.}
\author[L. Inno et al.]{
L. Inno,$^{1,2}$\thanks{E-mail: inno@mpia.de}
M.~A.~Urbaneja,$^{3}$
N.~Matsunaga,$^{4}$
G.~Bono,$^{5,2}$
M.~Nonino,$^{6}$
V.~P.~Debattista,$^{7}$\newauthor
M.~Sormani,$^{8}$
M.~Bergemann,$^{1}$
R.~da Silva,$^{9}$
B.~Lemasle,$^{10}$
M.~Romaniello,$^{11}$ 
H-W.~Rix$^{1}$
\\
$^{1}$Max-Planck Institute for Astronomy, D-69117, Heidelberg, Germany\\
$^{2}$INAF--OAR, via Frascati 33, I-00040 Monte Porzio Catone, Rome, Italy\\
$^{3}$Institut f\"ur Astro- un Teilchenphysik, Universit\"at Innsbruck, Technikerstr. 25/8, A-6020 Innsbruck, Austria\\
$^{4}$Department of Astronomy, School of Science, The University of Tokyo,7-3-1 Hongo, Bunkyo-ku, Tokyo 113-0033, Japan\\
$^{5}$Dipartimento di Fisica, Universit\`a di Roma Tor Vergata, via della Ricerca Scientifica 1, I-00133 Rome, Italy\\
$^{6}$INAF-Oss. Astr. di Trieste, Via Tiepolo 11, 34131 Trieste, Italy\\
$^{7}$Jeremiah Horrocks Institute, University of Central Lancashire, Preston PR1 2HE, UK\\
$^{8}$Universität Heidelberg, Zentrum für Astronomie, Institut für theoretische Astrophysik, Albert-Ueberle-Str. 2, D-69120 Heidelberg, Germany\\
$^{9}$Agenzia Spaziale Italiana, via del Politecnico snc, 00133 Rome, Italy\\
$^{10}$ Astronomisches Rechen-Institut, Zentrum für Astronomie der Universität Heidelberg, Mönchhofstr. 12-14, 69120, Heidelberg, Germany\\
$^{11}$European Southern Observatory, Karl-Schwarzschild-Str. 2,D-85748 Garching bei Munchen, Germany \\
}

\date{Accepted XXX. Received YYY; in original form ZZZ}

\pubyear{2018}

\begin{document}
\label{firstpage}
\pagerange{\pageref{firstpage}--\pageref{lastpage}}
\maketitle

\begin{abstract}
We report the discovery 
from our IRSF/SIRIUS Near-Infrared (NIR) variability survey 
of five new classical Cepheids located in the 
inner Galactic Disk, at longitude $l\simeq -40^{\circ}$. 
The new Cepheids are unique in probing
the kinematics and metallicity of young stars 
at the transition between the inner Disk and the
minor axis of the central Bar, where they are expected 
to be less affected by its dynamical influence.
This is also the first time that metallicity of Cepheids is estimated 
on the basis of medium-resolution ($R\sim3,000$) NIR spectra, 
and we validated our results with data in the literature, finding a minimal dependence 
on the adopted spectroscopic diagnostics. 
This result is very promising for using Cepheids as stellar proxy 
of the present-time chemical content of the obscured regions in the Disk.
We found that the three Cepheids within 8--10 kpc from us
have metallicities consistent with the mean radial metallicity gradient, and
 kinematics consistent with the Galactic rotation curve.
Instead, the closest ($\sim$4~kpc)/farthest ($\sim$12~kpc) 
Cepheids have significant negative/positive residuals, 
both in velocity and in iron content.
We discuss the possibility that such residuals are related 
to large-scale dynamical instabilities, 
induced by the bar/spiral-arm pattern, 
but the current sample is too limited to reach firm conclusions.
%


\end{abstract}




\section{Introduction}
A comprehensive empirical characterisation of the
present-time  Milky  Way  (MW) disk  is pivotal 
to pursue a full understanding of the formation history  of  our  Galaxy. 
While the structure of its central components, such as the Bar and the Bulge, have been recently unveiled by large-scale surveys,  
 their dynamical interplay and  their mixing with the underlying disk-component remains poorly understood.
The inner Disk, defined as the region within a Galactocentric distance ($R_{GC}$) from 4 to 7 kpc
\citep{feltzing13}, is the Galactic locus where all these components coexist and, likely, interact.
On the basis of stellar counts and photometric surveys, it is now well established that 
within the inner Disk lies the interface between the Disk, its inner spiral-arms
and the long Bar, with a half length of $\sim$5 kpc \citep[see e.g.][for a recent review on the Galactic Bar and Bulge]{bland-hawthorn16},
but it is still unclear e.g. where the Bars' ends and resonances are located \citep{benjamin05}, 
wether there is a central molecular ring \citep{heyer15} and which mechanism triggers the intense star-forming activity found e.g. in W43 \citep{bally10}, etc.

Moreover, the inner Disk is  expected to be strongly affected by the bar-driven dynamical instabilities \citep{rix13},
which impact both its kinematics and the chemical enrichment, 
and therefore the observed radial metallicity gradient.
Indeed, gas-phase metallicity gradients in barred external galaxies are systematically
shallower than the ones in un-barred galaxies, as supported by
empirical \citep[e.g.][]{zaritsky94,allard06}
and theoretical investigations \citep[e.g.][]{athanassoula92,friedli95,minchev13},
while the stellar-phase shows a higher degree of complexity \citep[e.g.][]{sanchez10}.
In the generally accepted inside-out infall scenario of the Galactic-disk
evolution \citep{chiappini97}, the radial migration 
induced by the Bar's instabilities produces a smoothing of
the metallicity content in the inner Disk. 
At the same time, the Bar's resonances 
allow a pile-up of fresh material from inflowing gas,
which produces hot-spots of star-formation activity, where local enrichment
can occur. The end-product of this mechanism is the presence of 
an azimuthal metallicity gradient \citep[e.g.][and reference therein]{khoperskov18}.

Despite the relevance of the role played by the inner Disk in the Galactic evolution, 
its kinematics and chemical content has not been empirically constrained yet, 
as a consequence of the significant dust extinction 
towards the Galactic plane, which makes observations challenging at low latitudes. 
There are a few studies based on bright hot stars \citep[e.g. OB stars,][]{daflon04},
while more investigations focused on the use of star clusters \citep[e.g.][]{jacobson16} to
trace the chemical pattern of the inner Galaxy.

In this context, Classical Cepheids are the best-suited stellar probes, 
as they are so luminous that can be  seen at distances larger than 8~kpc from us, 
even through substantive dust extinction,
and their individual distances and ages can be 
precisely determined on the basis of their pulsation periods.
They are very young stars \citep[ages $\lesssim$ 200 Myr][]{bono05},
but they have lower effective Temperature (T$_{\rm{eff}} \sim$~6,000K) with respect to  
to other stars of similar age and hence their spectra, rich in metal absorption lines, allow for
precise abundance determinations 
of many different chemical elements \citep[e.g.][ and references therein]{dasilva16}.

Indeed, works based on the chemical abundances 
of  Cepheids within 4--7~kpc from the Galactic Center,
have shown that the metallicity gradient steepens 
in the inner Disk \citep{andrievsky02,pedicelli09,luck11,genovali13},  
and it can reach super solar metallicity of $\sim$0.4~dex
at R$_{GC}\sim$2.5~kpc \citep{martin15,andrievsky16}.  
However, such investigations rely on only a few Cepheids
($\sim$5) at galactocentric radius of $\lesssim$5~kpc,
and thus at the transition with the Bar. 

In fact, the number of Cepheids currently known in inner regions of the Disk \citep[][Fig.~11]{metzger98b}
is still limited with respect to the ones found in the solar neighbourhood,
because variability surveys, traditionally performed in the optical bands, 
were strongly hampered by dust extinction towards the Galactic Center. 
Even the ongoing Gaia space mission will suffer 
from severe limitations in the Galactic plane, 
where its capability of detecting variables brighter than $\sim$20 mag (G-band), 
will translate to Galactocentric distances $\gtrsim$6~kpc 
due to extinction \citep{windmark11}.
Near-infrared (NIR) variability surveys have instead the capability 
to probe the innermost regions of the Galaxy by overcoming 
the thorny problems of the high extinction and the differential reddening.

In particular, our NIR survey towards the Galactic plane \citep{matsunaga11a}, 
started in 2007 at the Japanese Infrared Survey Facility (IRSF), 
pioneered the search for distant Cepheids 
in the NIR bands and led to the discovery of four new Cepheids located 
in the Nuclear Bulge \citep{matsunaga11a,matsunaga15}, 13
new Cepheids beyond the Galactic Center \citep{matsunaga16}
and three new Cepheids within R$_{GC} =3-5$~kpc in the Northern inner Disk \citep{tanioka17}.
At the same time, NIR data collected by the ESO public survey 
VVV \citep[VISTA survey of the Via Lactea][]{minniti10} also reported the discovery of 
 additional 24 Cepheids located beyond the GC \citep{dekany15a,dekany15b}.
 However, determining the chemical abundances of the Cepheids identified
 by such NIR surveys also requires the use of NIR spectroscopy,
 which remains challenging for dynamical supergiant stars like Cepheids.
 In fact, while abundance determination of Cepheids based on optical spectra 
is well established, the NIR regime is still in its infancy,
 and a very limited background study is available in the literature \citep[see e.g.][]{sasselov90,nardetto11}.

In this paper, we report the discovery of five new Cepheids located  
in the inner Disk, at longitudes $\sim -40$~degrees and close to the minor axis of the central Bar.
As a follow up of the photometric observations, we collected medium-resolution (R$\sim$3,000)
NIR spectra with ISAAC@VLT for all the new Cepheids 
and we were able to determine their kinematics and iron abundances.
The new Cepheids are the first ever detected in this region of 
the Disk and allow us to investigate the kinematics and chemical content 
of young stars in such crucial transition 
region. %

The paper is organised as follows. 
In Section~\ref{sec: phot}, we describe our photometric survey, 
the photometric data and data-reduction,
and we present the NIR light-curves of the new Cepheids,
 together with their distance estimates (Section~\ref{sec: dist}).
Then, we describe the spectroscopic data-set in Section~\ref{sec: spec},
 and the analysis techniques for optical and NIR spectra, 
 including the calibration of an homogenous metallicity abundance scale and the determination of the Cepheids' velocities. 
 We investigate the kinematics of the new Cepheids in the inner Disk in
 Section~\ref{sec: kin}, and their metallicity as a function of galactocentric radius (Section~\ref{ssec: rad})
 and azimuthal angle  (Section~\ref{sec: azit}).
Finally, in Section~\ref{sec: conc}, we summarise and discuss our results in the context of the Galactic disk evolution.


\section{Observation and detection of the new Cepheids}
\label{sec: phot}
We conducted a NIR survey in the Galactic plane towards 
the line of sights $l$ = -20$^{\circ}$,  -30$^{\circ}$ and  -40$^{\circ}$ 
with the IRSF 1.4m telescope and the SIRIUS camera, which takes images 
in the three NIR bands $J$ (1.25 $\mu$m), $H$  (1.64 $\mu$m) and $K_{\rm{S}}$ (2.14 $\mu$m) 
simultaneously \citep{nagashima99, nagayama03}.  
The instrumental field of view is about 7.7' $\times$ 7.7', with a pixel scale of 0.45"/pix.  
Between 2007 and 2009, thirty to forty observations have been performed 
for nine fields along the selected lines of sight.
For each direction, we carried out monitoring observations for a 20'$\times$ 20' area, 
covered by a 3$\times$3 IRSF/SIRIUS fields of view.
To obtain the photometry of the stars in these fields, 
we applied the same analysis presented in \citet{matsunaga09b}. 
After the pipeline reduction we carried out point-spread-function (PSF) fitting photometry 
using the DAOPHOT package in IRAF\footnote{IRAF -- Imaging Reduction and Analysis Facility-- is distributed by the National Optical Astronomy Observatory, which is operated by the Association of Universities for Research in Astronomy, Inc., under cooperative agreement with the National Science Foundation \url{http://iraf.noao.edu/}}.  
Among all the images collected in each filter, the one characterised by the best condition 
(sky transparency and seeing) was selected as a reference frame 
for all the other observations of the same field. 
The photometric results of this reference frame 
were compared with the 2 Micron All-sky Survey (2MASS) 
point source catalogue \citep{skrutskie06} in order to standardise the magnitudes. 
The reference frame has then been adopted to calibrate the remaining images of the same field. 
Thus, we created a master list of objects detected in each image, 
and we searched for their variability by adopting the same technique described in \citet{matsunaga13}. 
This technique identifies the stars for which the variation in magnitude 
is larger then three times the standard deviation computed on all the images 
of the same object in a given filter. 
Among the many variable stars discovered ($\sim$100, Matsunaga et al., in prep), 
we found five objects for which the light-curve shape is compatible with the ones of Classical Cepheids. 
These Cepheids are the first ones ever detected in the IV Galactic quadrant ($-90 \leqslant l \leqslant 0^{\circ}$)  and in the inner Disk (ID), 
and thus have been renamed ID-1, ID-2, ID-3, ID-4 and ID-5, in order of increasing Right Ascension. 


\begin{figure*}
\begin{center}
\includegraphics[width=0.90\textwidth]{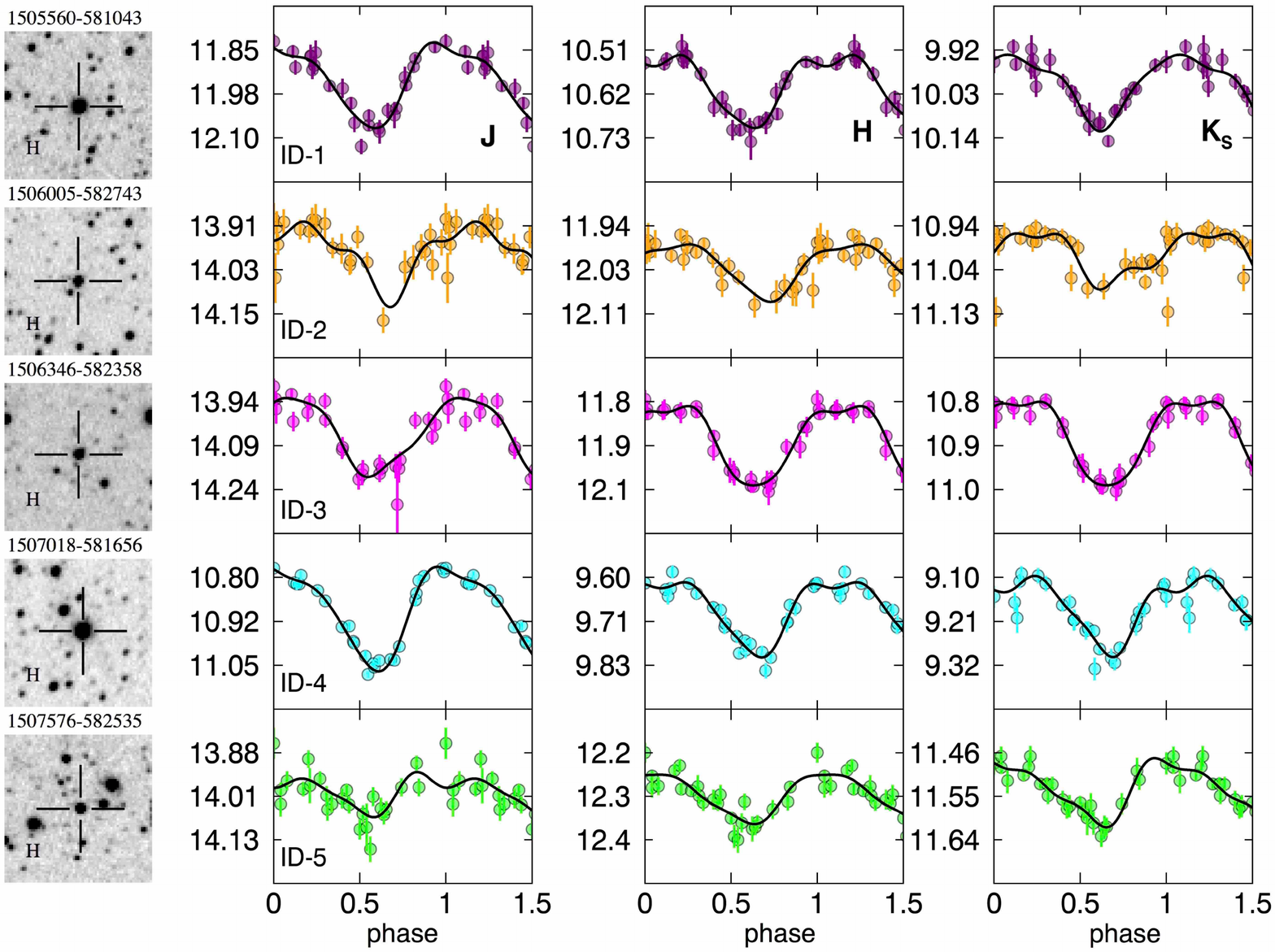}
\vspace*{-0.1truecm}
\caption{From top to bottom: observed light curves in the $J$,$H$ and $K_{\rm{S}}$ band for the five newly discovered Cepheids (ID1: purple; ID2: orange;
ID-3: magenta; ID-4: cyan; ID-5: lime) interpolated by the fourth order Fourier-series best-fit (black line).  
The error bars are the photometric error associated to the observed magnitude, while the typical $rms$ ranges from 0.05 (ID-5) to 0.08 mag (ID-3) in the $K_{\rm{S}}$ band}
\label{fig1}
\end{center}
\end{figure*}  

\begin{figure*}
\begin{center}
\includegraphics[width=0.90\textwidth]{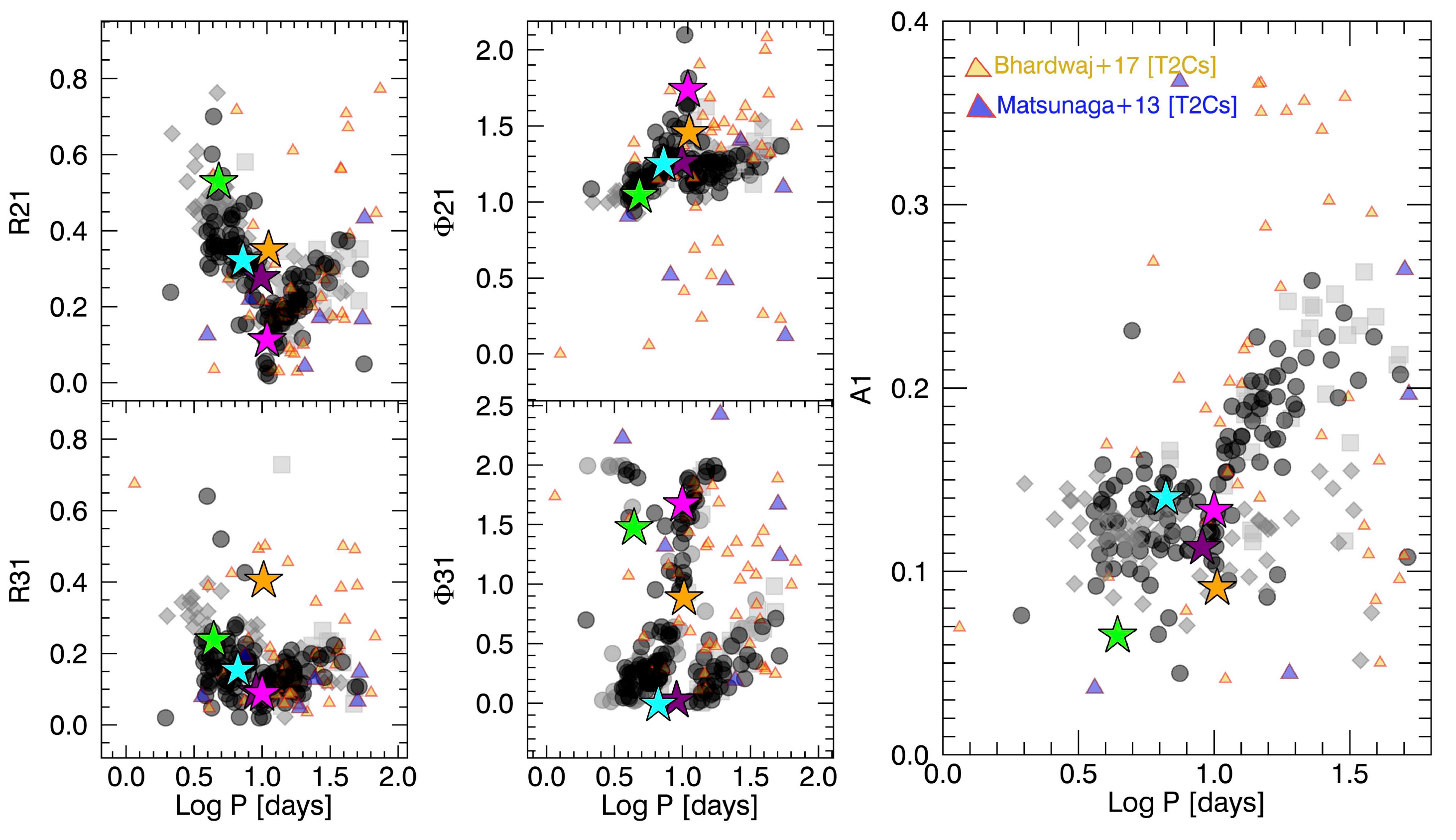}
\vspace*{-0.1truecm}
\caption{Fourier parameters of $J$-band light-curves for the newly discovered Cepheids (same colors as in Figure~\ref{fig1}) 
compared to the parameters of known Classical Cepheids in the Milky Way  \citep[][dark dots]{laney92,monson11},  
in the Large \citep[][grey squares]{persson04} and  
in the Small \citep[][grey diamonds]{inno15} Magellanic Clouds. 
We also include data for  Type~2 Cepheids in the MW \citep[][purple triangles]{matsunaga13} and in the Large Magellanic Cloud
\citep[][yellow triangles]{bhardwaj17}.
This comparison clearly show how similar the Fourier parameters 
of the newly identified variables are to the ones of other Classical Cepheids already reported in the literature}
\label{fig2}
\end{center}
\end{figure*}

%
%
\begin{table*}
	\centering
	\caption{Coordinates and Photometric data of the newly discovered Cepheids}
	\label{tab1}
	\begin{tabular}{lcccrrrr}
	\hline
  	Name & RA & Dec & Period &   $<J>$ &   $<H>$ &  $<K_{\rm{S}}>$  \\
	  	& [$h:m:s$] &	  [$d:m:s$]&	  [days]&	  [mag]&	 	  [mag]&	  [mag]\\
         	 \hline 
ID-1  [purple] &15:05:55.61 & -58:10:43.5 & 9.036 & 11.939 $\pm$ 0.003 & 10.596 $\pm$ 0.003 & 10.005 $\pm$ 0.004 \\
ID-2 [magenta] & 15:06:00.57 & -58:27:43.3 &10.26 & 13.985 $\pm$ 0.007 & 12.024 $\pm$ 0.005 & 11.007 $\pm$ 0.005  \\
ID-3 [orange] &15:06:34.67 & -58:23:58.1  & 9.996 & 14.045 $\pm$ 0.005 &  11.938 $\pm$ 0.004 &  10.873 $\pm$ 0.003  \\
ID-4 [cyan] &  15:07:01.83 & -58:16:56.1  &  6.644 & 10.891 $\pm$ 0.001 &  9.685 $\pm$ 0.001 &  9.178 $\pm$ 0.003  \\
 ID-5 [lime] &15:07:57.59 & -58:25 :5.7  & 4.407 & 13.991 $\pm$ 0.005 &  12.317 $\pm$ 0.005 & 11.533 $\pm$ 0.01  \\
  	 \hline 
	\end{tabular}
\end{table*}
%
We estimated the period $P$  of the new identified Cepheids 
by performing a fourth-order Fourier series fitting 
to the photometric data. 
The $J-$, $H-$ and $K_{\rm{S}}-$band light-curves of the new Cepheids 
together with their analytical fits are shown in Figure~\ref{fig1} (ID-1: purple, ID-2: light red; ID-3: magenta, ID-4: cyan, ID-5: lime)
and the resulting flux-averaged magnitudes in each band are listed in Table~\ref{tab1}. 
The error bars show the photometric errors, which range from 
$e_J\sim$0.01 mag (at $J \sim$11 mag) to $e_J\sim$0.1 mag (at $J \sim$16 mag). 
The $J$-band light-curves of the new Cepheids show indeed features typical of known Classical Cepheids 
with similar periods, such as e.g. the appearance of a bump on the rising branch related to the
so-called Hertzsprung Progression starting at period $\geqslant$7 days \citep{inno15}.
In order to properly quantify these proprieties, we determined the
Fourier parameters of the $J$-band light-curves: $A_1$, $R_{21}$,$R_{31}$,$\Phi_{21}$,$\Phi_{31}$ as defined by \citet{simon81}.
These parameters are listed in Table~\ref{tab2} and shown in Figure~\ref{fig2}, where they are compared to 
the ones of known Cepheids in the Milky Way \citep[][dark dots]{laney92,monson11},  
as well as in the Large \cite[][light grey squares]{persson04} and  
in the Small Magellanic Cloud \citep[][grey diamonds]{inno15}.
Unfortunately, extensive datasets of  $J$-band light-curves for Type~2 Cepheids
are not available in the literature, thus it is difficult to obtain
accurate Fourier parameters of these variables.
However, we included the available data for Type~2 
Cepheids towards the Galactic Center \citep[][purple triangles]{matsunaga13},
and in the Large Magellanic Cloud \citep[LMC,][yellow triangles]{bhardwaj17,macri15} .
Even though there is some marginal overlapping for some of the Fourier parameters between
population 1 and 2 Cepheids, when considered altogether, the parameters of 
the new variables are only compatible with the ones of known classical Cepheids, thus
further confirming our classification. 

%
%
 \begin{table}
	\centering
	\caption{Fourier Parameters for the $J$-band light-curves of the new Cepheids}
	\label{tab2}
	\begin{tabular}{lcccccc}
	\hline
  	Name & A$_1$ & R$_{21}$ &  R$_{31}$ &   $\Phi_{21}$ &  $\Phi_{31}$  \\
	  	&  [mag] &	 	 &	  &	        $\pi$&      $\pi$  \\
         	 \hline 
	ID-1  &  0.113 & 0.276 & 0.092 & 1.26 & 0.02 \\
	ID-2  &  0.091 & 0.349 & 0.404 & 1.45 & 0.88 \\
	ID-3  &  0.133 & 0.112 & 0.087 & 1.74 & 1.67\\
	ID-4  &  0.140 & 0.321& 0.153 & 1.25 & -0.01\\
	ID-5  &  0.065 & 0.528 & 0.237 & 1.04 & 1.47\\
  	 \hline 
	\end{tabular}
\end{table}

\subsection{The Cepheid Galactocentric distances}
\label{sec: dist}
\begin{figure}
\begin{center}
\includegraphics[width=\columnwidth]{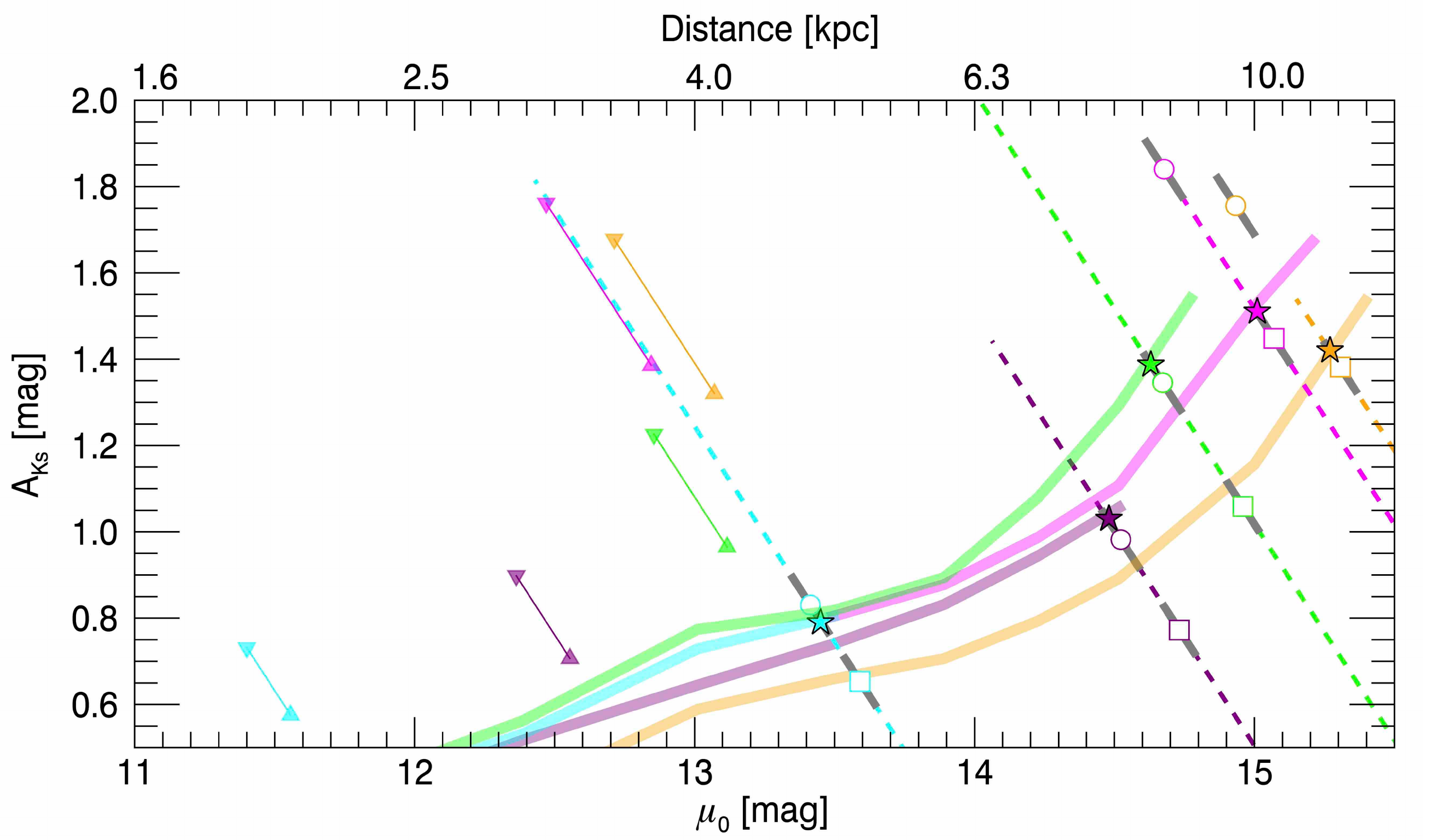}
\vspace*{-0.1truecm}
\caption{Absorption in the $K_{\rm{S}}$-band as a function of the true distance modulus 
from the $mwdust$ map \citep{bovy15} for the five new Cepheids 
(solid lines, ID-1: purple, ID-2: light red; ID-3: magenta, ID-4: cyan, ID-5: lime). 
The dashed lines are lines of constant $\mu_{K}$,
and their intersections with the solid lines, which are indicated by the stars, 
provide the estimate of the Cepheids' distances on the basis of the dust map.
The empty circles indicate the distances obtained by using the C89 extinction coefficients,
while the empty squares are the distances based on N06. These values indicate the minimum and 
maximum values that the Cepheids' distances can attain.  
Finally, the triangles show distances and extinction obtained by assuming that the new variables are
Type~2 Cepheids, by adopting the C89 (triangles up) or the N06 (triangles down) reddening law. 
These values are clearly inconsistent with the predicted absorption along the line-of-sight, 
thus further confirming our classification as classical Cepheids.
}
\label{fig3}
\end{center}
\end{figure}

The multi-wavelength observations of the new Cepheids allow for a simultaneous determination of their
distances and extinction by using Period-Luminosity (PL) relations in two different bands and 
by computing the corresponding selective extinction coefficients on the basis
of the assumed extinction law. However, recent works \citep[e.g.][]{nataf16, schlafly16, majaess16}  
have shown that the extinction law in the Milky Way is patchy and can vary as a
function of the line-of-sight, even in the NIR regime. Such uncertainty on the adopted
extinction law strongly impacts Cepheids' distance determinations, 
especially in highly obscured Galactic regions,
as extensively discussed e.g. in Matsunaga et al. (2018, subm). 
As described in \citet{tanioka17}, we can use the extinction laws by \citet[][hereinafter C89]{cardelli89} 
and by \citet[][hereinafter N06]{nishiyama06} to determine our best estimates of the Cepheids' distances
and the likely maximum uncertainties. In fact, the published values of the exponent $\alpha$ for the power law
used to model the extinction, mostly range between the two values determined by these two authors, namely
 1.61 (C89) and 1.99 (N06).
We use a similar approach here:
in order to determine the true distance modulus and extinction,
we adopt the PL relations in the $H-$ and $K_{\rm{S}}-$ bands
estimated by \citet{inno16} for the LMC Cepheids, 
calibrated by assuming a mean distance modulus to the LMC: 18.493 $\pm$ 0.008 $\pm$ 0.047 mag
\citep{pietrzynski13} and transformed into the IRSF photometric system \citep{kato07}.
In fact, the true distance modulus, defined as:
\begin{equation}
\mu_0=\mu_{K_{\rm{S}}}-A_{K_{\rm{S}}}, 
\end{equation}
can be derived on the basis of the apparent distance modulus:
$\mu_{K_{\rm{S}}}= <K_{\rm{S}}>- <M(K_{\rm{S}})>$, and the selective absorption in the $K_{\rm{S}}$-band:
$A_{K_{\rm{S}}}=A_{K_{\rm{S}}}/E(H-K_{\rm{S}}) \times (\mu_{H}-\mu_{K_{\rm{S}}})$,
by assuming first the coefficient $A_{K_{\rm{S}}}/E(H-K_{\rm{S}}) = 1.83$ by C89,
and then $A_{K_{\rm{S}}}/E(H-K_{\rm{S}}) = 1.44$ by N06.
The values obtained for the same Cepheids with the different extinction laws are 
listed in column 2 and 4 of Table~\ref{tab3} 
and they differ from 0.2 mag (ID-1,4) up to 0.4 mag (ID-2,3), or 16\% in distance, which is considered
as our \emph{conservative} systematic uncertainty.
However, in order to determine our best estimates,
instead of adopting the mean values,
we use a comparison with the distances determined
on the basis of a three-dimensional dust map.
In fact, \citet{drimmel03} computed such a map 
for the fourth Galactic Quadrant, where the new Cepheids are located,
and \citet[][hereinafter B16]{bovy15} incorporated
it  in a comprehensive Galactic extinction map. Thus, we used
the B16 $mwdust$ code to determine the amount of selective
absorption in the visual band for the five line-of-sights 
where the new Cepheids are located and at different distances. 
Note the dust map computed by
\citet{drimmel03} rely on the dust emission in the Far-Infrared 
as measured by the COBE/DIRBE mission,
on an assumed model for the Galactic structure,
and marginally on the choice of the extinction law,  
which is necessary to derive the opacity from 
the optical depth along the line of sight \citep[see Eq.~24 in][]{drimmel01}.
Thus, the map does not allow us to obtain a new determination
of the Cepheids' distances completely independent of the extinction law.
However, \citet{drimmel03} adopted the 
extinction law by \citet{rieke85} 
and internally validated their choice by comparing 
the extinction predicted with the one measured
for red-clump and OB stars in the inner and outer Galaxy.
Here, we adopt the coefficient by C89, in order to extrapolate
$A_{K_{\rm{S}}}$=$0.117 \times A_{V}$ in the IRSF photometric system, 
but the difference between the coefficients should be small (less then 1\%). 
This is necessary because \citet{rieke85} do not provide a general function for the extinction curve 
which could be extrapolated to different wavelengths, but they only provide $A_{K}$=0.112 $\times A_{V}$.
The values of  $A_{K_{\rm{S}}}$ obtained from the dust map
as a function of the true distance modulus is shown 
as a thick solid line in Figure~\ref{fig3} for the five line-of-sights 
where the Cepheids are located (same colors as in Figure~\ref{fig1}).
The lines $\mu_{K_{\rm{S}}}$=constant for each Cepheid are indicated by the dashed lines in Figure~\ref{fig3}, 
and their intersections with the solid lines of the corresponding colors, highlighted 
by the stars, determine the true distance modulus obtained on the basis of the extinction map (see Table~\ref{tab3}).
In Figure~\ref{fig3} we also show the distances derived on the basis of  the $H-$,$K_{\rm{S}}-$ bands PL relations, and 
assuming the C89 (empty circles) and the N06 (empty squares) coefficients, 
together with the associated errors, indicated by the grey error-bars.
Note that the distance values can only change according to the dashed lines,
and by assuming that the extinction law cannot be shallower then the C89 in the inner Galaxy,
we consider the empty circles as the minimum possible distance for the five new Cepheids.
Similarly, we cannot expect them to be any further of the distances indicated by the empty squares.

More specifically, the data plotted in this Figure show that for Cepheids: ID-1, ID-4 and ID-5, 
the use of C89 reddening law provide values that are closer to 
the prediction from the dust map, and thus we adopt them as as our best estimates.
For Cepheids ID-2 and ID-3, instead,  the N06 law gives
values more consistent with what is predicted by the map, and we decided to 
take them as the best estimates for these Cepheids.
The final adopted distances and their associated errors are given in the last column of Table~\ref{tab3}.
On the basis of these distances, and by assuming a $J$-band PL relation,
we can also obtain the best estimate of the extinction coefficient
$A_{K_{\rm{S}}}/ A_{J}$ for each of the Cepheids,
and we find that it is close to the C89 value (0.40) for two of them, while it attains 
slightly larger values than the one by N06 (0.33) for the others,
thus confirming that the slope of the extinction law in this direction is likely in between the two.

The coloured triangles show the values of extinction and distances
for new variables in the case they are Type~2 Cepheids.
Also in this case, we assumed the PL relations by \citet{bhardwaj17}
in the $H$ and $K_{\rm{S}}$ bands calibrated to the LMC distance,
and the coefficients first from C89 (triangles up) and N06 (triangles down).
The values of distances and extinction obtained 
are inconsistent with the absorption along the line-of-sight as predicted by the dust map.
Thus, this hypothesis can be discarded. 
However, Cepheids ID-4 and ID-5 have short periods,
thus they could be classical Cepheids but pulsating in the first overtone. 
In this case, they would be intrinsically brighter of $\sim$0.5 mag \citep[see e.g.][]{feast97,kovtyukh16},
and therefore $\sim$1 kpc further. However, first-overtone pulsators
are characterised by a different range of Fourier parameters and by smooth sinusoidal light-curves, without 
the bumps clearly visible on the rising branch of the light-curves in Figure~\ref{fig1}.


%
%
\begin{table*}%
	\centering
	\caption{Distances and extinction of the newly discovered Cepheids}
	\label{tab3}
	\begin{tabular}{lccccccrc}
	\hline
	  Name &  $\mu_{0,C89}^{(a)}$ &  $A_{Ks,C89}^{(a)}$ &  $\mu_{0,N06}^{(b)}$ &  $A_{Ks,N06}^{(b)}$ & $\mu_{0,B16}^{(c)}$ &  $A_{Ks,B16}^{(c)}$ & Distance$^{(d)}$ & $(A_{Ks}/A_{J})^{(e)} $\\
	       &	  [mag]&	  [mag]&	  [mag]&	  [mag]&	  mag]&	  [mag]&	    [kpc] & \\
          \hline 
 
ID-1  &  14.52 $\pm$  0.06  &    0.98 &  14.73  $\pm$ 0.05 &     0.77  &    14.48  &     1.0    &     7.9 $^{+0.8}_{-0.2}$ & 0.41 \\
ID-2  &  14.93  $\pm$ 0.06 &    1.76  &   15.31   $\pm$ 0.05 &    1.38   &    15.27  &     1.5    &   11.5$^{+0.3}_{-1.9}$ & 0.36 \\
ID-3   & 14.68  $\pm$ 0.06 &    1.84   &   15.07    $\pm$ 0.05 &   1.45 	&  15.01   &    1.5   &   10.5$^{+0.3}_{-1.8}$ &   0.36 \\
ID-4  &  13.42  $\pm$ 0.06 &    0.83   &  13.59    $\pm$ 0.15 &   0.65   &  13.45   &     0.8    &   4.8$^{+0.5}_{-0.1}$ & 0.37 \\
ID-5  &  14.67  $\pm$ 0.06  &   1.35   &  14.95    $\pm$ 0.15 &   1.06 	&  14.63   &     1.4    &   8.7$^{+1.3}_{-0.2}$  & 0.40 \\
     \hline 
	  \multicolumn{9}{l}{$^{(a)}$ Based on the extinction law by \citet{cardelli89} and  $H$-,$K_{\rm{S}}$-band PL relations.}\\	
	   \multicolumn{9}{l}{$^{(b)}$ Based on the extinction law by \citet{nishiyama06} and  $H$-,$K_{\rm{S}}$-band PL relations.}\\	
	   \multicolumn{9}{l}{$^{(c)}$ Based on the dust map by \citet{bovy15} and $K_{\rm{S}}$-band PL relation.}\\
	    \multicolumn{9}{l}{$^{(d)}$ The best distance estimates according to data shown in Figure~\ref{fig3} with the associated total (random and systematic) uncertainty.}\\
	     \multicolumn{9}{l}{$^{(e)}$ Extinction coefficients determined on the basis of our best distances.}\\
  	 \hline 
	\end{tabular}
\end{table*}

\begin{figure*}
\begin{center}
\includegraphics[width=0.70\textwidth]{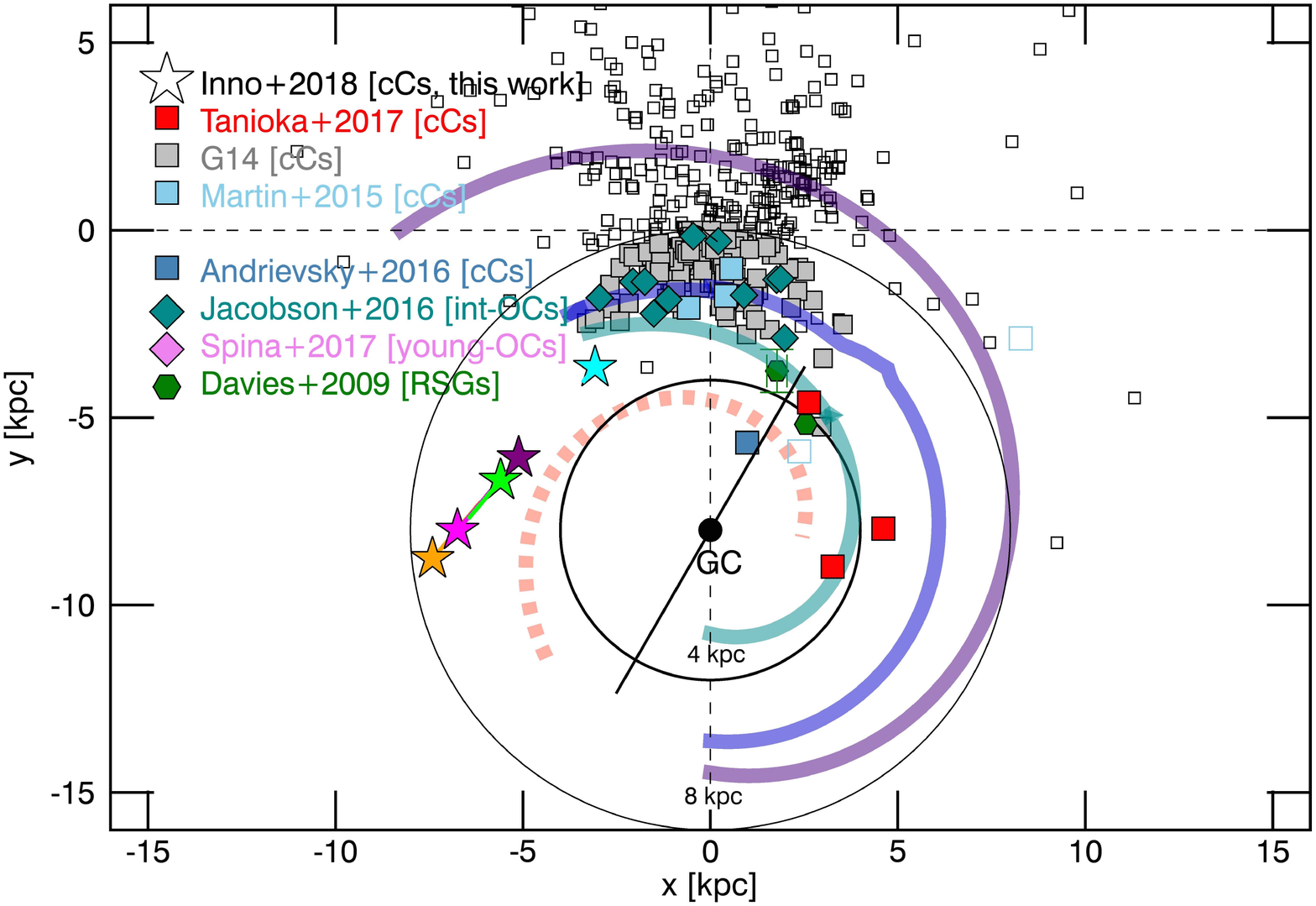}
\vspace*{-0.1truecm}
\caption{Position of the newly discovered Cepheids projected onto the Galactic plane  
(ID-1: purple, ID-2: light red; ID-3: magenta, ID-4: cyan, ID-5: lime). 
The squares show the position of currently known Galactic Cepheids
\citep{tanioka17,andrievsky16,martin15, genovali14}
while the blue rhombs indicate young and intermediate-age Open Clusters for which accurate abundances and positions
have been provided by the Gaia-ESO Survey \citep{jacobson16,spina17}.
The green hexagons shows the position of the two red supergiants clusters (RSGC1 and RSGC2) presented in \citet{davies09b}. 
Filled symbols indicate all objects located within 8 kpc from the Galactic Center and inside 250 pc above and below the Galactic Plane. 
The spiral-arm pattern of our Galaxy as recently revised by \citet{reid16} is also over plotted: 
the Scutum arm (dark cyan),  the Sagittarius arm (blue), the Perseus arm (indigo) and the Norma arm (red). 
The black lines shows circles at galactocentric radius of $R_{GC}$=4~kpc and of $R_{GC}$=8~kpc.
The position of the Sun (yellow dot), of the Galactic Center (black dot) and the central long Bar according to \citet{bland-hawthorn16} are also shown.}
\label{fig4}
\end{center}
\end{figure*}

The position of the new Cepheids, projected onto the Galactic plane 
is shown in Figure~\ref{fig4}.
We used a cylindrical coordinate system $(x,y,z)$ centred on the Sun position, 
which is located at a distance from the Galactic center:  
$R_{\odot}$=7.94$\pm$0.37$\pm$0.26 kpc \citep[][and references
therein]{groe08,matsunaga13}.
We also show the position of all the 
known Galactic Cepheids for which homogenous metallicity  
were estimated by our group\footnote{We removed BC~Aql from the sample, 
because a recent improved coverage of its optical light curve indicates it is
a Type~2 Cepheid (see the AAVSO website at \url{http://www.aavso.org/vsx/} for more details)}
 \citep[gray diamonds,][hereinafter G14]{genovali14},
 by \citet[][blue squares]{martin15} and by \citet[][ASAS181024--2049.6, blue diamond]{andrievsky16}.
Filled symbols indicate all the objects in the inner Disk, defined as
the region within 2.5 and 8~kpc from the Galactic Center $and$ within 250~pc from the Galactic plane. 
Note that we do not include here the Cepheids found beyond the Galactic Center by \citet{matsunaga16,dekany15a,dekany15b}
as there is no spectroscopic observations for them available in the literature, thus they are not relevant for
the present discussion.
 
In order to investigate possible correlations between the new Cepheids' position 
and the spiral arms of our Galaxy, 
we also plot their distribution according to \citet{reid16}, including 
the Scutum (dark-cyan line), the Sagittarius (violet line) and the Perseus  (indigo line)
arms. We also show the expected location of the Norma arm according to a 4-arm 
logarithmic spiral model in red, but the exact morphology of the spiral arms 
in the fourth Galactic quadrant is still very uncertain \citep{vallee17}. 
The position of the Red Supergiant clusters RSC1 and RSC2 discovered by \citet{davies09b} is also shown, 
since they are located at similar galactocentric distances with respect to the new Cepheids, 
but at positive $l$, therefore close to the edge of the Galactic bar.
Finally, the position of 5 young \citep[$\lesssim$ 100 Myr,][magenta diamonds]{spina17}  
and 12 intermediate-age  \citep[$\gtrsim$100 Myr,][dark-cyan diamonds]{jacobson16} 
Open Clusters (OCs) investigated by the Gaia-ESO survey is also shown for comparison. 

By using the Period-Age relation determined by \cite{bono05}, we can determine the individual ages 
of the new Cepheids and we found that three of them have ages of about 45 Myr, 
while ID-4 and ID-5 are about 20 Myr older. Thus, only the young OCs
and the RSCs have ages comparable to the ones of the Cepheids.


\section{The Spectroscopic Data-set}
\label{sec: spec}
During the semester March--September 2013, 
we collected NIR spectra with ISAAC at VLT for all the new Cepheids 
discovered in the inner Disk and for the Cepheid V367~Sct, 
which we adopt as a calibrating Cepheid to compare  
metallicity abundances based on NIR spectroscopy 
with the one derived from optical spectra. 
We adopted the medium-resolution (R$\sim$3,100) spectroscopic (SWS1\_MR) grism 
with the central wavelength $\lambda$=1.185 $\mu m$. 

\begin{figure*}
\begin{center}
\includegraphics[width=0.65\textwidth]{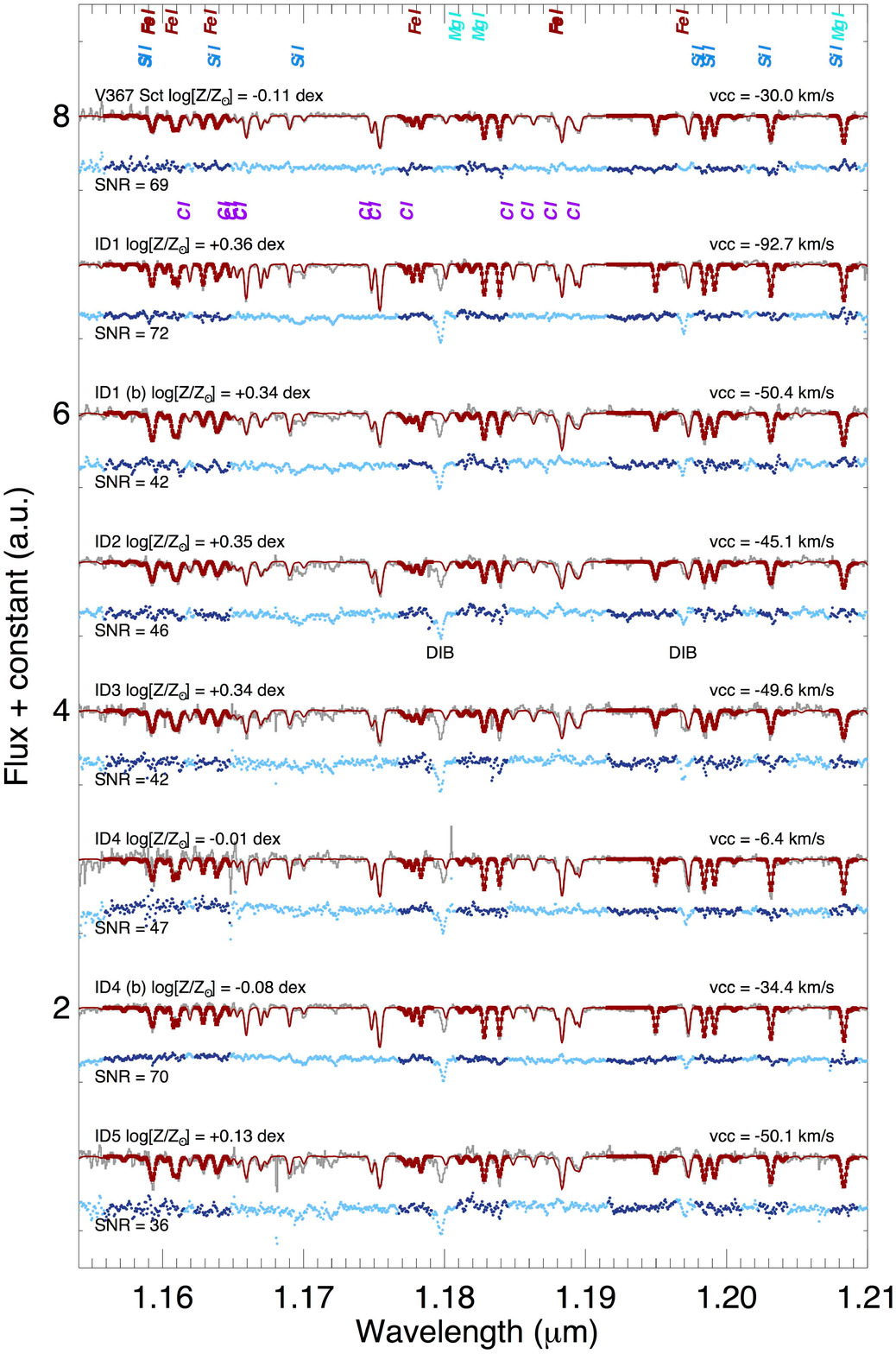}
\vspace*{-0.1truecm}
\caption{The analysed ISAAC spectra for the different targets. Some of the identified lines are labelled in the top (mainly Fe I, C I, Si I and Mg I). 
The SNR and the doppler velocity of the individual spectra is also given in the label.}
\label{fig5}
\end{center}
\end{figure*}

We have retrieved the raw files from the ESO Science Archive Facility, 
using the CalSelector option to associate the appropriate calibration files to the science ones. 
Due to the ABBA observing strategy, the sky subtraction has been done in the usual A-B/B-A way.
Spectra extraction on the single frame has been accomplished using \emph{apall}. 
The extracted spectra were then normalised via the \emph{continuum step} in IRAF. 
The wavelength solution has been directly derived from the telluric standards, using as a primary reference 
a template absorption spectrum obtained from the 
ESO software \emph{Skycalc}\footnote{\url{http://www.eso.org/observing/etc/bin/gen/form?INS.MODE=swspectr+INS.NAME=SKYCALC}}. 
The target spectra were then cross-correlated with the
wavelength calibrated telluric (in pixels) and shifted accordingly.
This allows both the telluric correction and wavelength calibration of the targets.
The reduced spectra were then corrected for atmospheric absorption features by adopting \emph{Molecfit}, 
which is a software developed by the Institute for Astro- and Particle Physics at the University of Innsbruck for ESO. 
This tool is based on synthetic transmission spectra which can then be directly fitted to the observed spectra. 
Finally, the spectral analysis has been performed by adopting a novel technique we recently developed. 
This technique is based on existing precomputed grids of synthetic LTE spectra based on the MARCS code. 
The stellar parameters and abundances are then obtained by a best-fitting of the synthetic to the observed spectra, 
where the fit is performed through a sophisticated automated analysis procedure based on Monte Carlo Markov Chain techniques.
The spectra reduced and corrected for the telluric lines are shown in Figure~\ref{fig5} 
 together with the identification of some of the lines used for the metallicity estimate, while the physical parameters measured are
 listed in Table~\ref{tab5}. Note that in the case of ID-1 and ID-4, we had two independent observations at two different pulsation phase,
 so we adopted the mean of the metallicity obtained from the two spectra, weighted for their SNR. 

 Some of the metal lines identified and used for performing the fit with the synthetic models are labelled in Figure~\ref{fig5}.
 Note that there are also several intense DIBs (Diffuse Interstellar bands) present in the spectra of the
 new highly-obscured Cepheids, which
 are not visible instead in the spectrum of the calibrator V365~Sct. 
 Excluding the most prominent one at 1.18 \citep{joblin90}, that is already known, the other DIBs are rarely observed, 
 because the line-of-sight extinction is not high enough. 
 This is the case of the structures centred around 1.17 microns and the one at 1.20 microns.

%
%
\begin{table*}%
	\centering
	\caption{Physical parameters, Iron abundances and kinematics of the newly discovered Cepheids}
	\label{tab4}
	\begin{tabular}{lrrccccccccc}
	\hline
  Name &
  Epoch&
  $\phi_{max}^J$&
   SNR &
  $T_{\rm{eff}}$&
  $\log g$&
  $v_t$&
  [Fe/H]&
  R$_G$&
       $V_{obs}^{(a)}$&
            $V_{LSR}^{(b)}$&
            $V_{gas}^{(c)}$
\\

	  &
	  MJD&
	  &
	  &
	  [$K$]&
	          &
	  [km s$^{-1}$]&
	  [dex]&
	    [kpc]&
	     [km s$^{-1}$]&
	       [km s$^{-1}$]&
	        [km s$^{-1}$]
	  \\

\hline

ID-1 &  6527.484 & 0.77 & 72 & 6738 & 2.12 & 6.4 &   0.36 $\pm$ 0.15  & 5.5 $\pm$ $^{+0.3} _{-0.1}$    &  -70 $\pm$ 1  & -60 $\pm$ 5&  -80 $\pm$ 10\\
ID-1 &  6407.688 & 0.51 & 42 & 6000  & 1.88 & 5.6  &   0.34 $\pm$ 0.15  &  $\textquotedbl$         &   -61  $\pm$ 1&  $\textquotedbl$  &    $\textquotedbl$  \\
ID-2 &  6407.723 & 0.51 & 46 & 6290 & 1.64  & 5.2 &  0.35 $\pm$ 0.15 & 7.5 $\pm$ $^{+0.2} _{-1.3} $  &   -56 $\pm$ 1 & -52 $\pm$ 5 &   -19 $\pm$ 10\\
ID-3 &  6407.773 & 0.83 & 42 &6227  & 1.64  & 5.2 &  0.34 $\pm$  0.15 &   6.8 $\pm$ $^{+0.2} _{-1.0} $   &   -60  $\pm$ 1 & -50 $\pm$ 5& -41 $\pm$ 10\\
ID-4 &  6401.898 & 0.88 & 47 & 6176 & 2.12  & 6.8 & -0.01 $\pm$  0.15 & 5.3 $\pm$ $^{+0.1} _{-0.1} $       & -19 $\pm$ 1 & -27  $\pm$ 5&  -89 $\pm$ 10\\
ID-4 &  6533.484 & 0.68 & 70 & 6210 & 1.64  & 5.6 & -0.08 $\pm$  0.15 &  $\textquotedbl$      & -13  $\pm$ 1 &  $\textquotedbl$   &     $\textquotedbl$ \\
ID-5 &  6407.832 & 0.26 & 36 & 6018 &  1.87 & 4.94 &    0.23 $\pm$ 0.15 & 5.8 $\pm$ $^{+0.7} _{-0.10}$  & -61 $\pm$ 1& -56 $\pm$ 5& -72 $\pm$ 10 \\
  	 \hline 
	  \multicolumn{11}{l}{$^{(a)}$ Heliocentric observed radial velocity}\\	
	   \multicolumn{11}{l}{$^{(b)}$ $\gamma$-velocity respect to the LSR}\\	
	   \multicolumn{11}{l}{$^{(c)}$ Velocity for gas particles at the Cepheids' location in the simulation by \citet{sormani17}}\\
	\end{tabular}
\end{table*}
 %
 %

\subsection{An homogenous Near-Infrared--Optical metallicity scale}
\label{subs: cal}
Cepheids have been widely adopted to investigate the Galactic disk radial metallicity gradient, 
by our group \citep{lemasle07,lemasle08, pedicelli09,genovali14, dasilva16} 
and in the literature \citep{yong06,sziladi07,luck11,luck11b,andrievsky16,martin15}. 
However, all these investigations rely on high-resolution $optical$ spectra, 
and therefore they are limited to bright and/or nearby Cepheids, 
while the metallicity gradient is mostly influenced by Cepheids located in the inner and outer regions of the disk. 
Because of the high amount of extinction, 
NIR spectroscopy is necessary to investigate the metallicity gradient in such critical regions. 
Unfortunately, while the method to determine the stellar parameters from optical spectra is very well known and tested,
NIR spectroscopy is still in its infancy and, in fact,  this is \emph{ the very first study 
in which Cepheids' parameters and chemical abundances
are obtained from NIR-medium-resolution spectra}.

\begin{figure}
\begin{center}
\includegraphics[width=0.95\columnwidth]{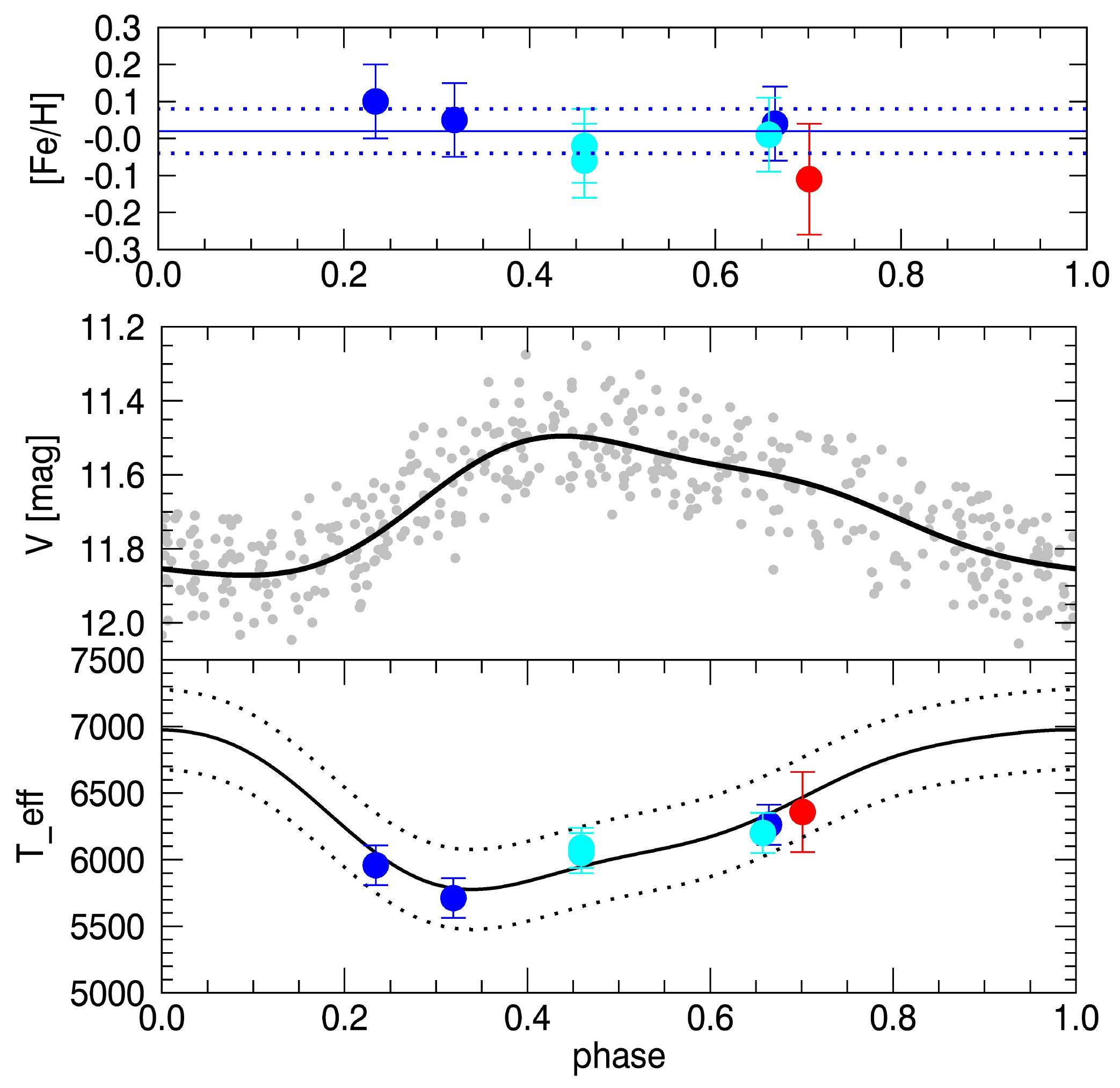}
\caption{Comparison of the calibrator Cepheid  metallicity and effective Temperature from optical and NIR spectra.
Top -- Iron-abundances obtained from the optical (UVES, blue dots; FEROS, cyan dots) and the NIR  (ISAAC, red dot) spectra
as a function of the pulsation phase at the time of observation.The blue solid line shows the mean value found on
the basis of the optical data, while the dashed lines indicate the $rms$. 
The value obtained with our analysis for the NIR spectrum is consistent within 1$\sigma$ with the ones in the literature.
Middle--Light curve in the $V$-band for the calibrating double-mode Cepheid V367~Sct from ASAS and the Fourier fit
to it. Bottom-- Effective Temperature obtained from the optical UVES  (blue dots) and FEROS (cyan dots) spectra
compared to the one obtained with ISAAC (red dot). The solid line is the mirrored light curve which is used 
to trace the variation of the effective Temperature due to the pulsation cycle,
while the dashed lines indicated the expected $rms$. The effective temperature
obtained by our new spectral analysis technique is in excellent agreement 
with the ones obtained by using a completely different approach \citep[see e.g.][]{ldr}
and wavelength regime (optical).
}
\label{fig6}
\end{center}
\end{figure}

Thus, we first need to ensure the compatibility 
of our results with the ones obtained by using the \emph{standard} 
approach as in \citet[][ and references therein]{genovali14}
To this purpose, we tested the metallicity abundances 
obtained from the ISAAC spectra of the calibrating Cepheid V367~Sct
against the ones obtained from four optical spectra collected at different epochs.
\citet{genovali14} analysed these spectra and found a mean abundance [Fe/H]=0.04 $\pm$ 0.08 dex, 
where the error is given by the $rms$ of the different measurements, 
and the typical uncertainty on individual determinations ranges from 0.13 to 0.18 dex. 
The individual estimates as a function of the pulsation phase 
are shown in the top panel of Figure~\ref{fig6}.
The mean value is indicated by the solid blue line, while the 
dashed lines show the $rms$ around the mean. 
On the basis of the ISAAC spectrum, 
we find [Fe/H]=-0.11 $\pm$ 0.15 dex (red dot in the top panel of Figure~\ref{fig6}), 
which agrees within 1$\sigma$ with the results from the optical data.
%
%

To further confirm the consistency of our NIR analysis,
we also compare the effective temperature obtained
from optical and NIR spectra of V367~Sct at the different observed phases.
The middle panel of Figure~\ref{fig6} shows the light curve in the $V$-band from 
ASAS\footnote{\url{http://www.astrouw.edu.pl/asas/?page=acvs}}. 
V367~Sct is a double mode Cepheids, folded at its fundamental period of 6.29307  days, 
therefore the large scatter in the photometric data.

We performed a Fourier fitting of the light curve and mirrored it
in order to qualitatively predict the shape of the temperature curve over the Cepheid pulsation cycle.
In particular, we normalised the curve, multiplied it for the expected temperature variation 
of a Cepheid along the pulsation cycle ($\sim$ 1,000 K) and
added the mean values of the temperature as determined from the spectra ($\sim$ 6,400 K). 
The new curve is plotted in the middle panel, where also the effective temperature 
measured on the optical (collected with UVES: blue dots; collected with FEROS: cyan dots) 
spectra \citep{genovali14,proxauf19} and on the ISAAC spectrum (red dot) are shown. 

This figure shows that the difference in temperature found in the optical and in the NIR regime
are consistent within the expected
change in temperature due to the pulsation mechanism.

%
%
\begin{table*}%
	\centering
	\caption{Physical parameters and Iron abundances of the calibrator Cepheid: V367~Sct}
	\label{tab5}
	\begin{tabular}{lcccccccc}
	\hline
  Epoch&
  $\phi_{max}^J$&
   SNR &
  $T_{\rm{eff}}$&
  $\log g$&
  $v_t$&
  [Fe/H]&
  Instrument &
  Resolution
  \\
    MJD &
	&
	  &
	  &
	  [$K$]&
	          &
	  [km s$^{-1}$]&
	  [dex]&
	  
	  \\
\hline 		
6405.0 & 0.70 & 100 & 6358  $\pm$ 300 & 1.9 & 6.3   &  -0.11 $\pm$ 0.15 &  ISAAC & 3,000$^{(a)}$   \\
6175.1 &  0.24 & 100 &  5959 $\pm$ 84 & 1.2  & 4.2 &  0.10 $\pm$  0.10 & UVES & 38,000$^{(b)}$  \\
4709.6 &  0.32 & 98 &  5712 $\pm$  93 & 1.3  & 3.4 &  0.05 $\pm$  0.10 & UVES & 38,000$^{(b)}$ \\
6184.0 &  0.67 & 98 &  6262  $\pm$ 196 & 1.0 & 3.4 &  0.04 $\pm$  0.10 & UVES & 38,000$^{(b)}$ \\
3156.8  & 0.46 &  150 & 6089 $\pm$ 107 & 2.0 & 4.9 &  -0.06 $\pm$  0.10 &  FEROS & 48,000$^{(b)}$ \\
3156.9  & 0.46 &  151 & 6111 $\pm$ 135 & 1.7 & 4.9 &  -0.02 $\pm$  0.10 & FEROS & 48,000$^{(c)}$ \\
3157.8 &  0.66 & 146 &  6212 $\pm$ 150 & 1.8 & 3.8 &  0.01 $\pm$  0.10 & FEROS & 48,000$^{(c)}$\\

  	 \hline 
	  \multicolumn{9}{l}{$^{(a)}$ This work.}\\	
	   \multicolumn{9}{l}{$^{(b)}$ Spectra analysed  in \citet{genovali14} and \citet{proxauf19}.}\\
	   \multicolumn{9}{l}{$^{(c)}$ Spectra analysed in \citet{kovtyukh16} and \citet{proxauf19}.}\\
	\end{tabular}
\end{table*}


\subsection{Cepheids's center-of-mass velocity from single-epoch spectroscopic data}
\label{sec: vel}

By comparing the observed to the synthetic spectra in the model grid
we obtained the instantaneous radial velocities of the Cepheids, which includes 
the barycentric velocity of the star (the so-called $\gamma$-velocity) and the 
pulsation velocity projected along the line of sight. 
Thus, we need to correct the observed velocity for the pulsation velocity
at the observed phase.
To this purpose, we computed new radial-velocity templates which allow us to
perform such correction \citep[e.g. see also: ][]{inno15,matsunaga15}.
We constructed velocity curve templates 
for Galactic Cepheids with period similar to the new ones,
and in particular for the two period ranges: $4 \lesssim P \lesssim 7$ (template-1) 
and  $9 \lesssim P \lesssim 10.5$ (template-2). 
Thus, we adopted the $K_{\rm{S}}$-band light curves 
and the radial velocities curves of 20 (template-1) and 8 (template-2) nearby Cepheids,  
compiled by \cite{storm11a}, to calibrate such templates as described in the following. 
We fitted a 7th-order Fourier series to both the light and radial-velocity curves, 
and computed the peak-to-peak amplitude of the fit.
The amplitude in the  $K_{\rm{S}}$-band,$\Delta_{Ks}$, is tightly correlated
with the velocity amplitude, $\Delta_{RV}$, 
and we calibrated the relations: 
\begin{eqnarray}
\label{eq: cs}
\Delta_{RV}/\Delta_{Ks} = 160 \pm 2.95~ {\rm km s^{-1} mag^{-1}, [template-1]}\nonumber \\
\Delta_{RV}/\Delta_{Ks} = 180 \pm 4.19~ {\rm km s^{-1} mag^{-1}, [template-2]}
\end{eqnarray}
which allow us to predict $\Delta_{RV}$
from the $\Delta_{Ks}$ of the new Cepheids.
Note that the quoted error is the dispersion around the mean  
$\Delta_{RV}$/$\Delta_{Ks}$ relation. 
In the case of the new Cepheids ID-1 and ID-4,
we collected ISAAC spectra at two different pulsation phases,
and the measured velocities are shown 
in Figure~\ref{fig4}, together with the radial-velocity template
adopted (template-2 for ID-1, top panel; template-1 for ID-4, bottom panel).
The final $\gamma$-velocities obtained are indicated by the dashed line,
while the shadowed area delimitates the error interval.
In fact, the uncertainty on the final estimate is given by the sum in quadrature of the 
error on the measurement of the doppler velocity (1 kms$^{-1}$),
the dispersion given in Equation~\ref{eq: cs}, 
and the scatter around the templates rescaled by the amplitude ($\sim$0.1$\times \Delta_{RV}$ kms$^{-1}$).  
%
\begin{figure}
\begin{center}
\includegraphics[width=\columnwidth]{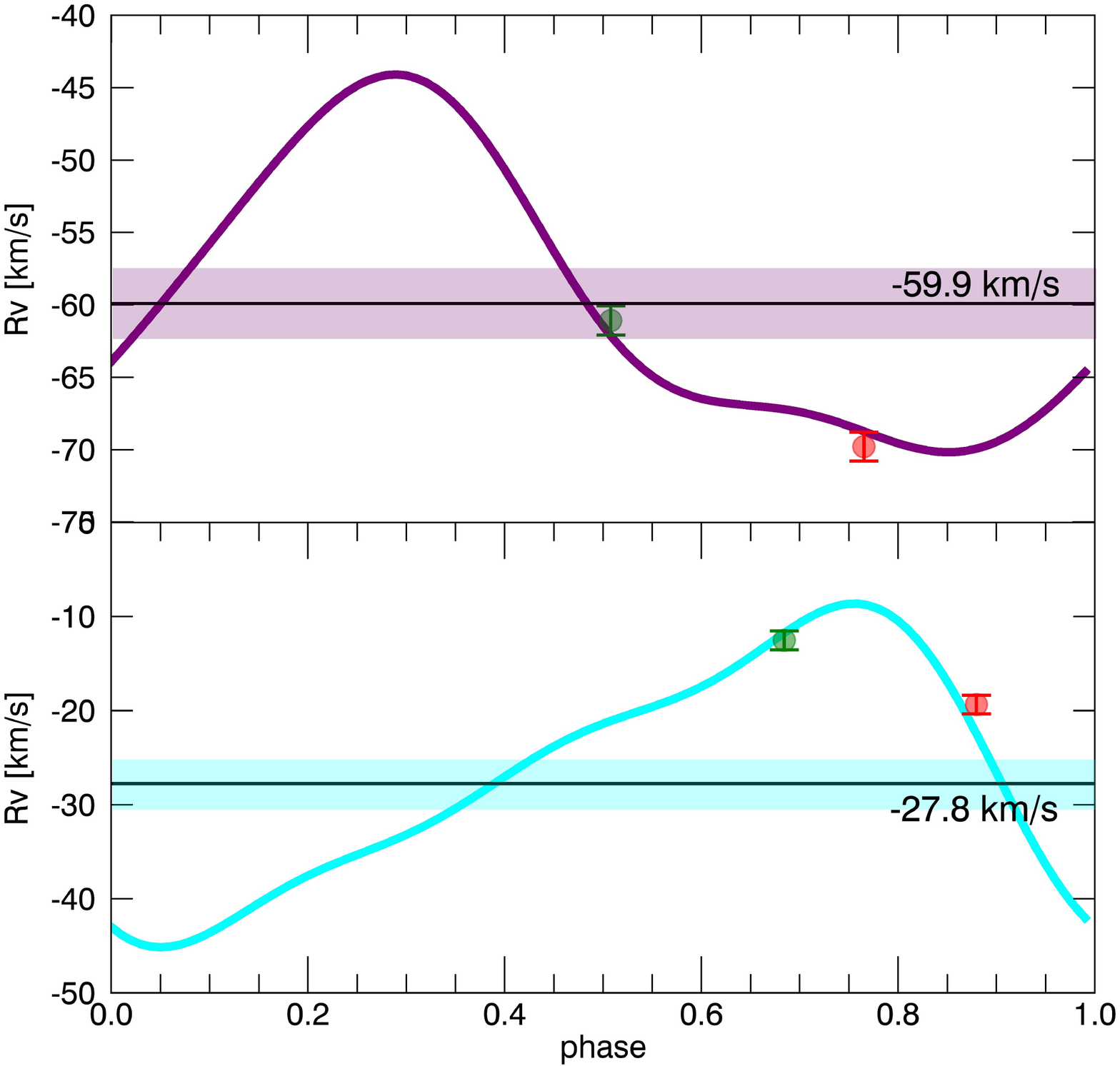}
\vspace*{-0.1truecm}
\caption{Radial-velocity templates applied to the new Cepheids ID-1 
(top panel, purple solid line: template-2, periods ranging from  9 to 10.5 days) 
and ID-4 (bottom panel, cyan solid line: template-1, periods ranging from  4 to 7 days) 
for which spectroscopic observations at two different phases were available. 
The dots show the observed heliocentric velocity of the new Cepheids at the phase of the first (red) and second (green) 
observations (see Table~\ref{tab3}).  
The solid black lines give the resulting mean radial-velocity, 
which corresponds to the $\gamma$-velocity of the Cepheids, while the shadowed area delimitate the $\pm$5 kms$^{-1}$ uncertainty. }
\label{fig7}
\end{center}
\end{figure}
%
Within the period ranges we selected,
the Cepheids show very similar variations in light and velocity,
and we can use the templates described above to predict 
the pulsation amplitude for all the new Cepheids. 
The resulting $\gamma$-velocities in the Local Standard of Rest (LSR), 
$V_{LSR}$, are listed in Table~\ref{tab4} and shown in Figure~\ref{fig8}.
The complete set of radial-velocity  templates for classical Cepheids
will be given in a forthcoming paper (Inno et al. in prep).

\section{The kinematics of young stars in the Inner Galaxy} 
\label{sec: kin}

\begin{figure}
\begin{center}
\includegraphics[width=\columnwidth]{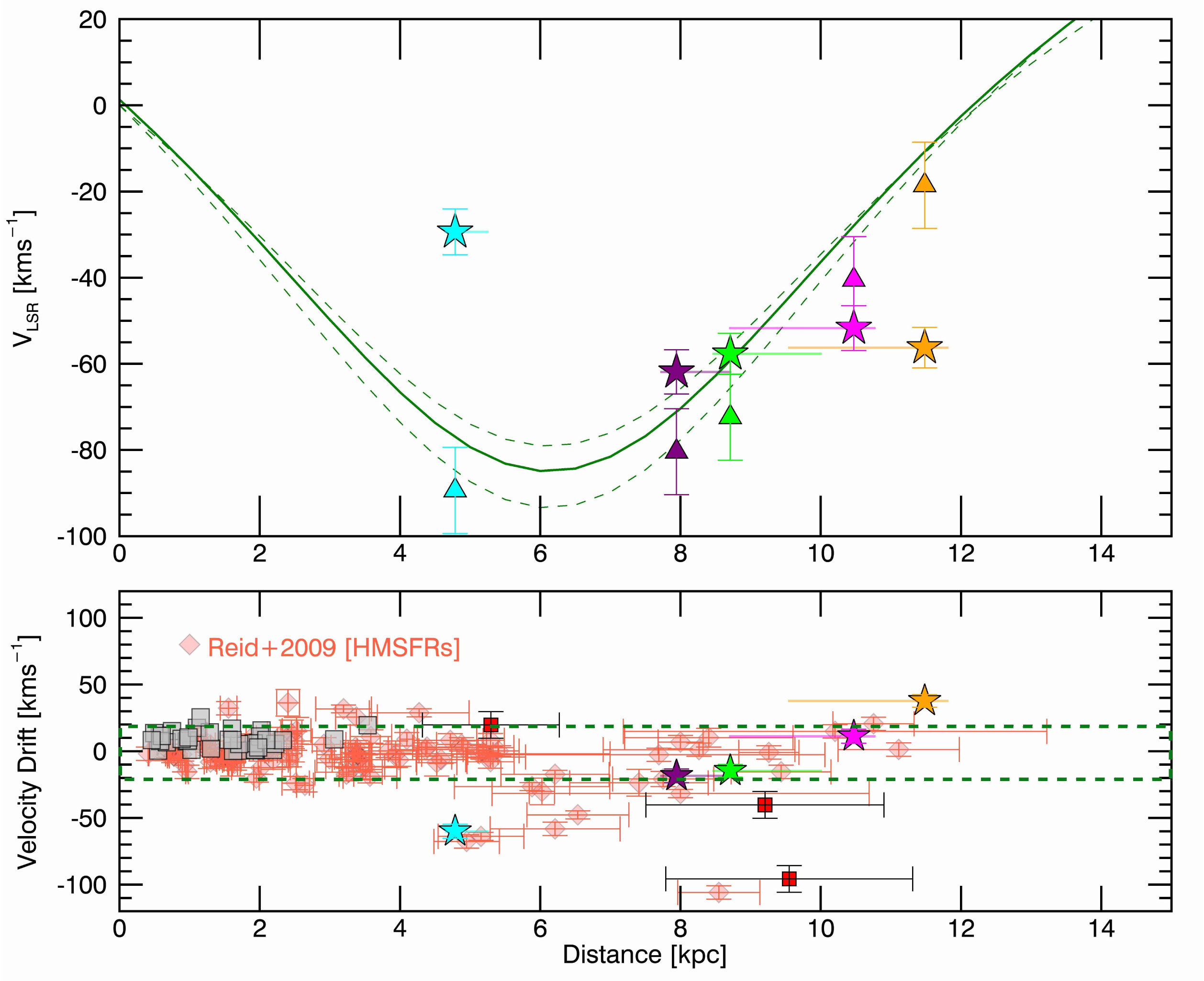}
\vspace*{-0.5truecm}
\caption{Top--
Line of sight velocity of the new Cepheids with respect to the LSR (coloured star marks) 
as a function of distance, compared to velocity predicted according to the Galactic rotation curve of  \citet[][solid green line]{reid14}. 
The dashed lines indicate a variation of $\pm$10 kms$^{-1}$ for the circular velocity of the Sun: V$_\odot$=240 kms$^{-1}$.  
Only two (ID-1, purple and ID-5, lime) out of the five new objects seem to have velocity consistent with the expected rotation, 
while the other three are either rotating slower (ID-4, cyan) or faster than expected.
The coloured triangles indicate the expected $V_{LSR}$ of the gas on the basis of the dynamical simulation by
\citet{sormani17} and are similar to those predicted from the the circular rotation curve.
Bottom-- Velocity drifts of the new Cepheids as a function of distance, compared to the 
ones of Cepheids in G14 with measured $\gamma$ velocity by \citet[][gray squares]{storm11a}, 
three Cepheids from \citet[][red squares]{tanioka17}, and the HMSFRs by \citet{reid14}. 
The green dashed lines indicate a difference of $\pm$20 kms$^{-1}$ from the predicted velocity. 
The new Cepheids have velocity drifts which seem to be proportional to the distance, 
thus indicating a systematically different velocity pattern. 
The HMSFRs at similar distances seem to follow a similar trend.
} 
\label{fig8}
\end{center}
\end{figure}

We compared the kinematics of the new Cepheids
with the one predicted by adopting the  rotation curve empirically calibrated by \citet{reid09},  which assumes
a solar distance $R_{\sun}$=8.0 kpc, and a circular velocity of the Sun 
$\Theta_{\sun}$=240 kms$^{-1}$ (solid thick line).
Figure~\ref{fig8} shows the expected velocity as a function of the distance $d$ from us,
together with the measured $V_{LSR}$ of the new Cepheids. 
Only  ID-1, ID-5 and marginally ID-3 show a kinematics consistent with the Disk rotation, 
while the others show significant differences.

Of course, an error on the estimated distances 
for the new Cepheids would also produce artificial velocity drifts. 
However, the Cepheid ID-4 should be at least $\sim$3~kpc closer to us or $\sim$2~kpc farther
in order to match the expected $V_{LSR}$, which would  
imply an error on the estimated distance larger than 3$\sigma$.  
This is quite unlikely, especially for this Cepheid, which is less obscured 
with respect to Cepheids ID-1 and ID-5, which have instead velocities consistent with the curve.  
Note that even in the case that the Cepheid ID-4 is pulsating in the first-overtone
mode, the velocity difference would still be observed, as the star would be located 
$\sim$1 kpc farther from us, where its expected velocity would be even larger. 

Similarly, an error on the $V_{LSR}$ would also produce the observed velocity drifts. 
The dominant source of uncertainty in the computed Cepheids' velocities is 
the correction for pulsation effects. 
However, it has been shown in the literature that Cepheids' pulsation-velocity 
amplitudes are $\sim$5--30 kms$^{-1}$\citep{storm11a,storm11b},
with possible modulations effects due to the rotation or presence of a companion  
ranging from several hundred ms$^{-1}$ to a few kms$^{-1}$ \citep{anderson16}.
The difference found here in the $V_{LSR}$ is significantly larger ($\sim$50 kms$^{-1}$). 
Once again, the velocity differences observed cannot be due to just measurements' errors. 
Moreover, we already found similar drifts for two Cepheids in the inner Disk 
but located at positive Galactic longitudes \citep{tanioka17} 
and concluded that either the Galactic rotation is slower in such regions,
or the kinematics of the Cepheids is affected by the dynamical instabilities of the Bar. 
%
%

Here, we find that such drifts are proportional to the distance,
and the disturbance caused by the presence of the Bar cannot account for them.
In order to check this, we used the simulations of  \citet{sormani17}, 
which model the gas flow in a Galactic barred potential 
and are able to reproduce the main observed characteristics 
of the atomic and molecular gas kinematics obtained through 
radio-frequency observations in the bar region ($l$<|30| deg).
These velocities are plotted as triangles of the same colors of the Cepheids
to which they refer to, and they are very similar to
the ones obtained on the basis of the purely circular rotation.
Therefore, the dynamical influence of the Bar is already
small at the distances where the Cepheids are located.

The bottom panel of Figure~\ref{fig8} 
shows the velocity drifts as a function of the heliocentric distance for the new Cepheids,
the three Cepheids discussed in \citet{tanioka17}, the Cepheids with
velocity estimates in \citet{storm11a} and the  High Massive Star Forming Regions (HMSFRs) 
used by \citet{reid09} to determine the Galactic rotation located in the inner Disk. 
Note that negative residuals at positive longitudes correspond to a faster rotation,
  while they correspond to a slower rotation at negative longitudes.
To avoid confusion, we define the velocity drift as the absolute difference 
between the observed radial velocity along the line-of-sight and the one
predicted on the basis of the pure circular rotation, and then we associate to it 
a positive sign at negative longitudes,
and a negative sign for positive longitudes. 
With this definition, a star with a positive value of the velocity drift is moving
faster than predicted from pure circular motion, while a star with a negative drift is moving slower
The drifts systematically increase for increasing distance in the case of the
new Cepheids, of two of the \citet{tanioka17} sample and of the HMSFRs at similar distances.  
This evidence indicates that such velocity pattern cannot be attributed to peculiar kinematics of the Cepheids,
but must be related to a large scale effect, such as e.g. the presence of the spiral-arms 
at these locations or of bar/spiral-arms interactions,
as in the cases of external spiral galaxies \citep[see e.g.][]{shetty07,beuther17}.

\begin{figure}
\begin{center}
\includegraphics[width=\columnwidth]{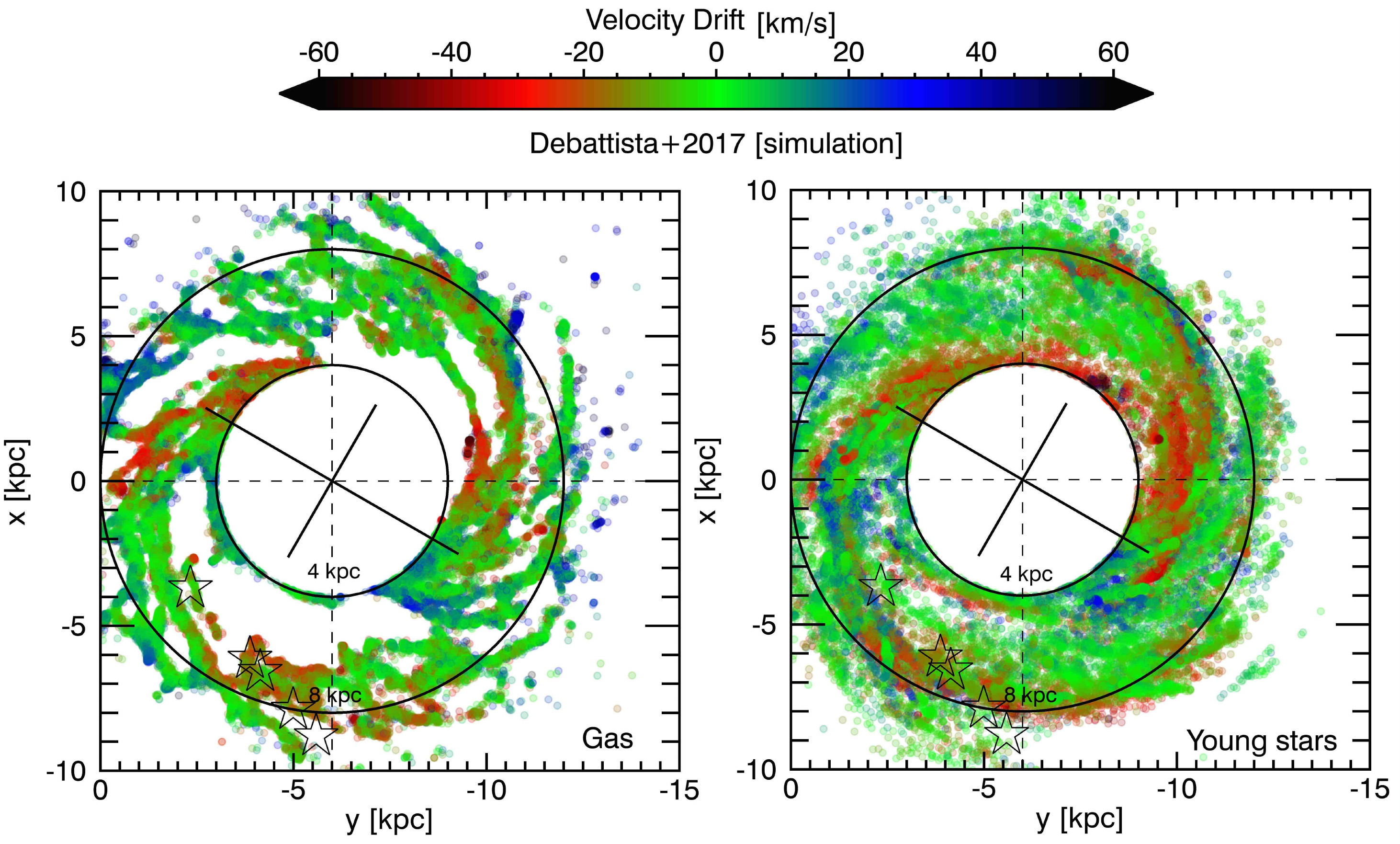}
\vspace*{-0.5truecm}
\caption{Deviation from circular orbits for gas (left panel) and young stars (age $\lesssim$ 300 Myr, right panel) 
in the simulation by \citet{debattista17}.
The central region (R$_{GC}\lesssim$4~kpc)
 is masked as it is dominated by the presence of the Bar, whose
 major and minor axis are also indicated.  
The color map indicates the
 velocity drifts as indicated by the colorbar.
Red stars are moving slower than expected, while blue stars are moving faster. 
The black empty stars indicates the positions corresponding to the new Cepheids. 
} 
\label{fig8b}
\end{center}
\end{figure}

To explore at least qualitatively this scenario, 
we show in Figure~\ref{fig8b} the distribution of cold gas (left panel) 
and young stars (age $\lesssim$0.3 Gyr, right panel) 
in the inner Disk, simulated by \citet[][see their Section~4]{debattista17},
and color-coded by the deviation from the expected rotation velocity.
While in the simulation by \citet{sormani17} the gas moves in an externally 
imposed potential in which the only non-axisymmetric component is the bar,
the simulation by \citet{debattista17} is self-consistent, with the potential obtained 
from a time-dependent N-body simulation in which spiral arms 
and other non-axisymmetric dynamical features are present in addition to the bar. 
Moreover, we rescaled the simulation in order to
match the physical size and inclination of the Milky Way central bar,
and the rotation velocity at the Sun location (8 kpc) of 240 kms$^{-1}$.
Thus, we computed the radial velocity V$_{LSR}$ along the line-of-sight
for each stellar particle in the simulated data, and then we subtracted  from it the
 value predicted by using the rotation curve by \citet{reid09}, as we did for the observed Cepheids.
 Moreover, we masked out the central region dominated by the Bar.
In fact, in the inner 4 kpc the deviation we computed has hardly a 
physical meaning, as the orbits are known to be not-circular due to the presence of the bar. 
Summarising, the right panel of Figure~\ref{fig8b} shows the position of stars 
that are moving either slower (red, negative drifts) or faster (blue, positive drifts) than expected. 
Such red and blue stars are not distributed uniformly
in the plane, but they seem to trace the edges of spiral-arm-like features.
This becomes even more evident by comparing their distribution to the one of the gas 
in the left panel of the same Figure. In fact,  the gas seems 
to move systematically slower at the leading edges and faster 
at the trailing ones of these spiral-arm-like features, 
which is consistent with recent observational results \citep{hunt16,meidt17,baba18}. 
This behaviour is probably related to the spiral-arm resonances, which allow
stars to spend enough time close to the arm itself, and to be initially decelerated by the 
local potential of the arm, and then accelerated when the arm has overtaken them.
The presence of non-circular orbits of stars due to such resonances in the inner Galaxy has been 
indeed already suggested by \citet{lepine11}
in order to explain the distribution and kinematics of molecular carbon monosulphides.
The same resonance could be also responsible for the observed drifts of the new Cepheids,
which have velocities consistent with the ones predicted by \citet{lepine11} for these non-circular orbits. 
Even though neither the available observed data or the simulated data allow us to completely 
characterise the orbits of the stars, it seems plausible that the drifts measured for the new Cepheids
is produced by the presence of the same or similar resonances, 
and that the systematic trend shown in the bottom panel of Figure~\ref{fig8} 
is related to the location of the stars across one or more spiral arms.

\section{Metallicity gradients in the inner Disk} 
\label{sec: grad}
\subsection{The radial gradient} 
\label{ssec: rad}
\begin{figure}
\begin{center}
\includegraphics[width=\columnwidth]{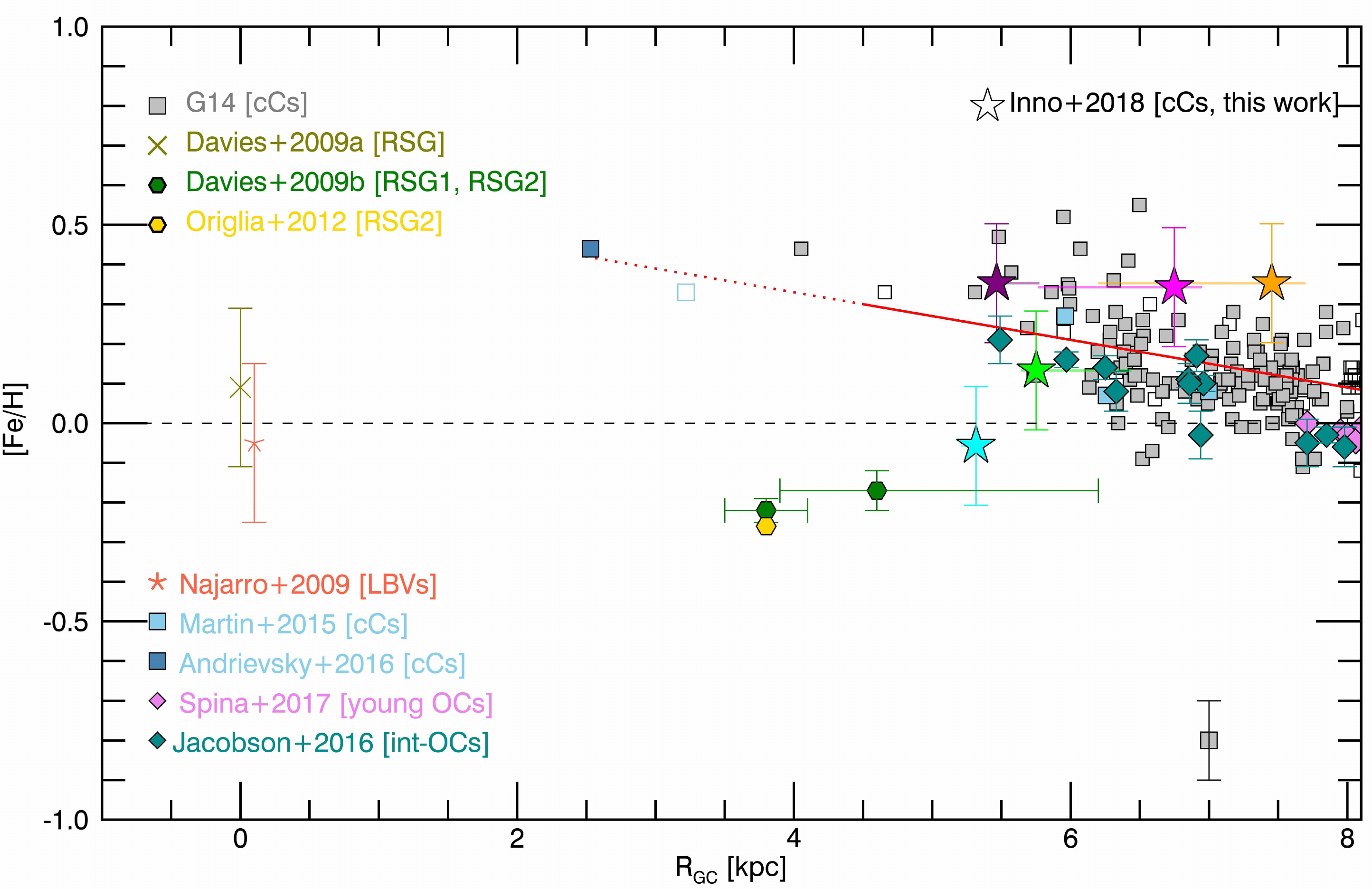}
\vspace*{-0.1truecm}
\caption{Iron abundances of of the new Cepheids (colored stars) versus their Galactocentric radius. 
The red line shows the metallicity gradient measured by G14 (solid line) and its extrapolation (dashed line)
towards smaller galactocentric radii, which can be compared
to the abundances of the Cepheids investigated by \citet{martin15} and \citet{andrievsky16}.
The hexagons mark the iron abundances of the two red supergiants clusters (RGSC1 and RGSC2) 
close to the bar's edge measured by \citet[][green]{davies09b} and by \citet[][yellow]{origlia13}, 
while the olive cross shows the metallicity of the RGSC 
in the Galactic center \citep{davies09a}. 
The red asterisk with its error Bar shows the metallicity 
range of luminous blue variables (LBVs) in the Quintuplet cluster \citep{najarro09}.
Finally, the violet diamonds show the metallicity of young (Age $\lesssim$ 100 Myr)  OCs measured by \citet{spina17}.
while the black triangle indicates the abundance for the calibrating Cepheid V367~Sct 
as estimated from the ISAAC spectrum.
 }
\label{fig9}
\end{center}
\end{figure}

In order to investigate the radial metallicity gradient in the inner Disk, 
we plotted in Fig.~\ref{fig9} the iron abundances of the newly discovered Cepheids 
as a function of the Galactocentric radius compared to the abundances of
all the other stellar tracers presented in Figure~\ref{fig4}:
the two red supergiants clusters (RGSC1 and RGSC2), 
as measured by \citet{davies09b} and by \citet{origlia13}. 
We also included the metallicity range covered 
by luminous blue variables (LBVs) in the Quintuplet cluster \citep{najarro09}. 
Note that all the iron abundances have been rescaled to the same solar abundance adopted in Section~\ref{sec: spec}.

The iron content of the new Cepheids is on average compatible (within the error bars)
with the one predicted by adopting a linear metallicity gradient on a wide range of Galactocentric
distances R$_G \sim$4--19 kpc, thus suggesting a homogenous chemical enrichment history,
at least on the timescale of Cepheids typical lifetime ($\sim$200 Myr).

However, one of the new Cepheids  (ID-4) has solar metallicity, 
and contributes at increasing the dispersion around the linear metallicity
gradient in the inner Disk.  In fact, given that 
the dispersion around the linear gradient reported in the literature 
is $\sigma$([Fe/H])$_{6.5  <{\rm R_{GC}/kpc} \lesssim 13 }\sim$0.09~dex, 
the Cepheid ID-4 shows a 2$\sigma$ discrepancy. 
This finding, together with the results for the RGSCs, indicates that 
the young stellar population in the inner part of the Galaxy 
cover a broad range of iron abundances: -0.3 dex $\lesssim$[Fe/H]$\lesssim$ 0.5 dex.
A  very similar metallicity range, i.e. -0.23 dex $\lesssim$ [Fe/H] $\lesssim$0.22 dex, 
is also spanned by luminous cool stars within 30 pc of the Galactic center, 
as found by \citet{cunha07}.
Since the extrapolation of the linear metallicity gradient to $R_{GC}\sim$0 would lead to significantly
super-solar abundance in the Galactic center, a flattening of the inner Disk gradient would be necessary to explain
the metallicity composition of these stars.
The metallicity content of the new Cepheids seems to suggest instead that the chemical composition of 
the inner part of the disk is characterised by a high degree of inhomogeneity.

\subsection{The azimuthal gradient} 
\label{sec: azit}
In order to investigate if the metallicity inhomogeneity is related to a more complex
spatial structure due e.g. to the presence 
of the Bar, we compare the new Cepheids' iron content with the 
metallicity of the other Cepheids in the inner Disk as a function of the azimuthal angle about the Galactic Center.
We performed this comparison following the argument by \citet{davies09b}. 
They suggest that objects in the inner Disk at negative $l$ tend to have super-solar abundances, 
while metal-poor RSG clusters are located at positive $l$. 
They argue that this evidence is well justified by the inside-out infall scenario,
when accounting for the dynamical effects due to the Bar's instabilities, 
since the edge of the Bar is towards positive Galactic longitudes (see Fig.~\ref{fig3}). 
This would reflect into an azimuthal metallicity gradient in the inner Disk. 

\begin{figure}
\begin{center}
\includegraphics[width=0.99\columnwidth]{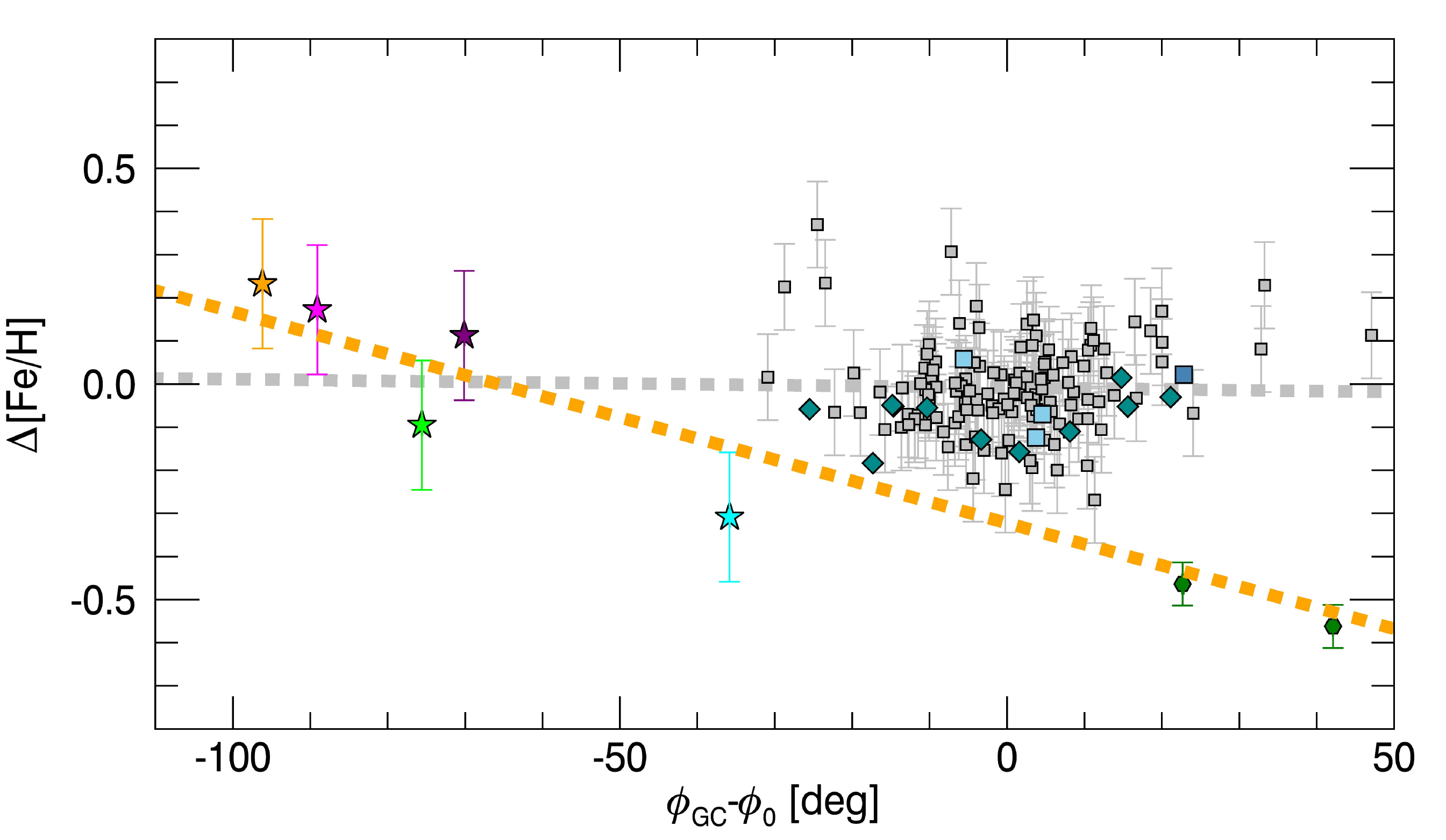}
\vspace*{-0.5truecm}
\caption{Residual with respect to the radial metallicity gradient as a function
of the azimuthal angle about the Galactic Center for the five new Cepheids
 and other stellar tracers in the inner Disk from the literature
(symbols are the same as in Figure~\ref{fig9}). We also show
a linear fit to all the Cepheids' data, which is indicated by the grey dashed line,
and it is consistent with a slope equal to zero (-0.000 $\pm$ 0.001).
The orange dashed line shows the best-fit to the metallicity residual restricted only to the new Cepheids
and the RSGCs, with a slope of -0.0057 $\pm$ 0.0008, thus indicating a negative gradient.
}
\label{fig10}
\end{center}
\end{figure}

We plot in Figure.~\ref{fig10} the residuals from the radial metallicity gradient of Cepheids 
as a function of their azimuthal angle, rotated of $\phi_0$=30$^{\circ}$,
i.e. the inclination angle of the Bar respect to the $x$-axis.

If we consider the sample of Cepheids in the inner Disk altogether,
the data do not seem to support
the presence of an azimuthal metallicity gradient, and indeed if we try to fit
a linear relation, we find a slope consistent with zero (grey dashed line in Figure.~\ref{fig10}).   
However, if we take into account only the newly discovered Cepheids and the 
RSGCs, we find a negative gradient significant at $7\sigma$ level.
A possible explanation of this result is that the RSGCs and the 
new Cepheids belong to the same spiral arm, the Scutum arm,
 weather the other Cepheids belongs to 
(several) different arms overlapping along the same Galactocentric direction. 
However, arm--inter-arm metallicity variations are expected to be of the order of $\sim$0.02 dex \citep{bovy14}, 
thus quite smaller than the trend we find.
Concluding, results based on Figure.~\ref{fig10} 
are difficult to interpret, as they are strongly biased by the incompleteness
of the Cepheid sample.  
In order to obtain more quantitative and conclusive constraints
on the metallicity distribution in the azimuthal directions,
a more extensive sample of Galactic Cepheids 
with a well defined selection function is needed.

\section{Conclusions}
\label{sec: conc}

We report the discovery of five Cepheids located in the fourth quadrant of the Galactic disk.
These Cepheids are unique in that they allow us to probe the transition between the inner Disk and the
central regions of the Galaxy along the minor axis of the central Bar,
where they are expected to be be less affected by its dynamical influences. 
We investigated the kinematics and iron abundance of the new Cepheids on the basis
of NIR medium-resolution spectra collected with ISAAC@VLT.  
We built specific radial-velocity-curve template in order to obtain
the line-of-sight systemic velocity of the Cepheids from the observed radial velocity. 
We also performed a careful comparison 
between abundances determined on the basis of optical high-resolution spectra
and NIR medium-resolution spectrum of the calibrating Cepheid V367~Sct. 
We found that the iron abundance estimates are consistent within the error bars, 
thus suggesting a negligible dependence on the wavelength regime and analysis techniques.
Summarising, we determined accurate distances, velocities and metallicities for the
newly discovered Cepheids, even though they are located in highly obscured regions
of the Galaxy.

In this paper, we show that the kinematics and the metallicity of these Cepheids are
different from the ones of already known Galactic Cepheids in the inner Disk. In fact, 
$a$) they move
slower then expected if located at small galactocentric radii and
faster if they are beyond the Galactic center;
and $b$) they are more metal rich at larger
then at smaller azimuthal angles. 
However, given the little number of new Cepheids identified and the patchy, inhomogeneous selection
function which characterises the entire sample of known Cepheids in the Milky Way,
any interpretation of such findings must be
taken with caution.
Nonetheless, our analysis show that
the kinematics and metallicity content of young stars in the inner Disk
is strongly influenced by dynamical instabilities induced by the bar/spiral-arm-pattern.

We used the simulation by \citet{sormani17}  to determine
the motion of the gas at the location of the Cepheids.
We found that the velocities predicted by the model are very similar to
 the ones predicted on the basis of the simple rotation curve
within 10 km/s, which is the uncertainty associated to the model.
We made the same comparison also for the three Cepheids presented in \citet{tanioka17},
and we found similar results. 

This indicates that the newly identified Cepheids are located too far with respect to the regions
where the Bar becomes dominant, thus bar-induced non-circular motions
alone cannot account for the velocity drifts observed. 
However, such dynamical instabilities can be induced by any non axis-symmetric potential,
also due e.g. to spiral arms, or even by bar/spiral-arm interactions. 
Recent results \citep[e.g][]{bobylev15, baba18} show that young stars 
have statistically significant velocity offset depending 
on their location either at the leading or trailing part
of the spiral arm. Such velocity variations are expected
to have amplitude of $\sim$6-10 kms$^{-1}$ with respect to the Galactic rotation.
These value are still 2--3 times smaller than the drifts we measured
for the Cepheids in the inner Disk, but they refer to the average perturbation, 
while the drifts can be larger for specific group of stars.

Summarising, by comparing results from the simulations of \citet{debattista17}
and \citet{sormani17} in Figure~\ref{fig8}~and \ref{fig8b} 
we can conclude that $i$) the bar cannot be the direct cause of the observed drifts; 
$ii$) drifts with amplitudes of $\pm$60 kms$^{-1}$ can be induced both in the gas and young stars 
by the presence of non-axisymmetric dynamical perturbations such 
as spiral arms outside the bar region ($R_{GC}$>4~kpc).
Moreover, the simulation by \citet{debattista17} shows that stars at the leading edges
of the spiral arms are mostly decelerated, while the ones at the trailing edges 
are accelerated, thus producing a systematic trend for velocity drifts
across the arms.
Finally, the observed deviations from the simple Galactic 
rotation also imply that the stars are moving in non-circular orbits, 
which are produced at dynamical resonances. 
Even though we cannot determine the new Cepheids orbits,
it is plausible that they are moving in boxy-shaped orbits
similar to the ones identified by \citet{lepine11} in the inner Galaxy.

The bar-like structure at the center of the Galaxy 
is also a key ingredient for Galactic evolution models
in order to explain the high star-formation-rate found in the Nuclear Bulge, 
since it is responsible for dragging gas and molecular clouds from the inner Disk 
into the center of the Galaxy \citep{athanassoula92, kim11}. 
This scenario implies that the metallicity distribution of the inner Disk 
should be similar to the one found in the Nuclear Bulge. 
Current findings show that the metallicity dispersion of the new Cepheids is indeed 
consistent with the metallicity of LBVs in the Quintuplet cluster by \citet{najarro09} 
and RSGCs in the Galactic Center by \citet{davies09a}.
Moreover, recent measurements by \citet{ryde15} for evolved stars in the NB (M giants) 
also show a similar large metallicity range, i.e. -0.13 dex$\lesssim$[Fe/H]$\lesssim$0.29 dex, 
with a good overlap with the metallicity range covered by the new Cepheids, i.e. -0.06 dex$\lesssim$[Fe/H]$\lesssim$ 0.33 dex. 

Concluding, the broad metallicity range covered by the new Cepheids suggest that 
 young and evolved stars in the two innermost 
Galactic regions with ongoing star formation activity, i.e. 
the inner Disk and the Nuclear Bulge,
have similar iron abundances, thus supporting a correlation between
their chemical enrichment histories.
On the other side, the absence of a clear azimuthal metallicity gradient rules out the 
dynamical effects of the Bar's instabilities as the cause of the large observed metallicity spread.
Thus, this dispersion seems then to be an intrinsic feature of the chemical composition of the Galactic inner Disk,
suggesting that its metallicity distribution cannot be modelled by using one-dimensional gradients.

This work shows that our understanding of the 
central regions of the Galaxy is still severely biased 
by the limited number of tracers available. 
Deep photometric surveys, especially in the NIR, 
together with NIR follow-up spectroscopy 
are necessary to achieve
more firms conclusions on the topic.
By succeeding in performing the  spectroscopic 
analysis of NIR spectra for pulsating supergiant stars
we contribute to make one step farther in this direction. 

\section*{Acknowledgements}
Based on observations collected at the European Organisation for Astronomical Research 
in the Southern Hemisphere under ESO programme: 290.D-5114(A), PI: L. Inno.
This work was supported by Sonderforschungsbereich SFB 881
"The Milky Way System" (subproject A3, A5, B1, B2 and B8) of the German Research Foundation (DFG).
NM is grateful to Grant-in-Aid (KAKENHI, No. 26287028) from
  the Japan Society for the Promotion of Science (JSPS).
VPD is supported by STFC Consolidated grant no. ST/M000877/1.
The star-forming simulation used in this paper was run
at the High Performance Computing Facility of the University of Central Lancashire.
We express our thanks to Shogo Nishiyama, Nagisa Oi, and Hirofumi Hatano
who collected a part of the IRSF photometric data for our targets.
We also warmly thank E. Valenti, for her help and many useful suggestions 
while serving as support astronomer for the ISAAC program 290.D-5114(A) PI: L. Inno,
as well as S. Meidt and O. Gerhard for the nice 
discussions on galaxies dynamics that helped us in improving the manuscript.

\bibliographystyle{mnras}
\bibliography{inner_disk} 

%
%
%
%
%
%

\bsp	
\label{lastpage}
\end{document}